\documentclass[twocolumn,9pt,superscriptaddress,a4paper]{quantumarticle}
\pdfoutput=1
\usepackage{amsmath}
\usepackage{braket}
\usepackage{graphicx}
\usepackage[colorlinks,citecolor=blue,linkcolor=blue]{hyperref}
\usepackage{array}
\usepackage{epsfig}
\usepackage{amssymb} 
\usepackage{dsfont}
\usepackage{times,txfonts}
\usepackage{soul}
\usepackage{xcolor}
\usepackage[numbers]{natbib}

\newcommand{\chng}[1]{\textcolor{black}{#1}}

\begin{document}

\title{Speeding up quantum annealing with engineered dephasing}
\author{Mykolas Sveistrys}
\affiliation{Fachbereich Physik and Dahlem Center for Complex Quantum Systems,
Freie Universität Berlin, Arnimallee 14, 14195 Berlin, Germany}
\author{Josias Langbehn}
\affiliation{Fachbereich Physik and Dahlem Center for Complex Quantum Systems,
Freie Universität Berlin, Arnimallee 14, 14195 Berlin, Germany}
\author{Rapha\"el Menu}
\affiliation{Theoretische Physik, Universität des Saarlandes, D-66123 Saarbrücken, Germany}
\author{Steve Campbell}
\affiliation{School of Physics, University College Dublin, Belfield, Dublin 4, Ireland}
\affiliation{Centre for Quantum Engineering, Science, and Technology, University College Dublin, Dublin 4, Ireland}
\affiliation{Fachbereich Physik and Dahlem Center for Complex Quantum Systems,
Freie Universität Berlin, Arnimallee 14, 14195 Berlin, Germany}
\author{Giovanna Morigi}
\affiliation{Theoretische Physik, Universität des Saarlandes, D-66123 Saarbrücken, Germany}
\author{Christiane P. Koch}
\email{christiane.koch@fu-berlin.de}
\affiliation{Fachbereich Physik and Dahlem Center for Complex Quantum Systems,
Freie Universität Berlin, Arnimallee 14, 14195 Berlin, Germany}

\date{09.04.2025}

\begin{abstract}
    Building on the insight that engineered noise, specifically, engineered dephasing can enhance the adiabaticity of controlled quantum dynamics, we investigate how a dephasing-generating coupling to an auxiliary quantum system affects quantum annealing protocols. By calculating the exact reduced system dynamics, we show how this coupling
    enhances the system's adiabaticity solely through a coherent mechanism - an effective energy rescaling.
    We show that it can lead to an annealing speedup linearly proportional to the strength of the coupling. 
    We discuss the experimental feasibility of the protocols, and investigate the trade-off between fidelity and implementability by examining two modified versions with fewer types of required physical couplings. 
\end{abstract}

\maketitle

\section{Introduction}

\label{sec:intro}
Quantum reservoir engineering has emerged as a promising paradigm for implementing robust quantum dynamics~\cite{Arimondo1996, ciraczollerreservoir1996, kraus2008reservoir, diehl2008reservoir, kienzler2015reservoir}. 
It can be summarised as controlling a quantum system by coupling it to a reservoir that is designed to induce some desired form of dissipation. The dissipative dynamics usually attracts the system to a steady state and, in many cases, the goal of reservoir engineering is precisely to obtain a desirable steady state. This state can be pure and have further interesting features, for example be highly entangled or belong to a non-trivial quantum phase, in the language of many-body physics~\cite{Verstraete:2008, kraus2008reservoir, diehl2008reservoir}.

In the spirit of quantum reservoir engineering, it has been suggested to realise the desired dissipation with a quantum measurement apparatus~\cite{pielawa2007generation, BurgarthPRL2007, raghunandan2020initialization}. Instead of directly engineering decay channels onto the system of interest, auxiliary quantum systems that interact with it---so called meters---are added. These meters can interact in a continuous way or have their state reset periodically, the latter case corresponding to collision models~\cite{CampbellEPL, CiccarelloReview}. Couplings with auxiliary quantum systems or meters provide significantly more flexibility in designing the decay processes since interactions involving more than two parties, albeit challenging, can feasibly be engineered, whereas ``natural" non-locality is rather rare. It follows that couplings to auxiliary degrees of freedom, as in measurement-induced dynamics or collision models, can improve the robustness and possibly the speed of adiabatic dynamics~\cite{raphael,King:2024}. 

A particularly relevant area of application are quantum annealing (QA) protocols, where realising an adiabatic dynamics is essential \cite{aqc}.
These protocols have shown great promise in solving combinatorial optimization problems and simulating quantum many-body systems. In the former category, QA has been used for traffic flow optimization~\cite{traffic, DeffnerTraffic2024}, scheduling~\cite{scheduling}, portfolio optimization~\cite{portfolio_opt} and materials design~\cite{materials}, among other problems. In the latter category, QA has allowed for observing phase transitions of spin glasses~\cite{phase_transitions}, topological phenomena~\cite{topological}, and quantum critical dynamics~\cite{critical}. 
While there is numerical evidence for speedups exhibited by QA for specific problems~\cite{scaling_advantage1, scaling_advantage2}, the case for a general problem-independent speedup is more disputed~\cite{qa_review, embedding_scaling}. QA, in the context of optimization, hopes to converge to a solution faster than its classical counterpart, simulated annealing, through the means of quantum tunneling~\cite{santoro2002theory, kadowaki1998quantum}. Nonetheless, it is a heuristic method without a proven speedup 
compared to classical optimization methods, which has resulted in several proposals to speed up annealing protocols~\cite{passarelli2022optimal, passarelli2023counterdiabatic, seki2012quantum}.

The biggest challenge in speeding up QA is diabatic transitions to excited states. The adiabatic theorem puts a fundamental limit to the annealing speed as a function of the energy gap between the ground state and other excited states~\cite{aqc}. Moreover, the time dependence of the gap sizes are generally unknown {\it a priori} such that one cannot simply slow down around regions where energy gaps are smaller. \chng{Some promising approaches to tackle these issues have been developed, e.g. engineering a site-specific dissipation~\cite{Najafabadi2023} or adapting the state preparation scheme with suitable preconditioning terms~\cite{CuginiPRR2025}}. Alternatively, provided the dynamics are known (at least approximately) methods based on shortcuts-to-adiabaticity (STA)~\cite{sta_review} allow for correcting diabatic transitions. They include using counterdiabatic driving~\cite{berry_cd} which can be combined with optimal control~\cite{delCampoPRA2014, cold}, invariant-based engineering~\cite{invariant_engineering}, \chng{and geometric considerations~\cite{Shingu2025}} to mimic the result of slow, adiabatic dynamics on, in principle arbitrarily, shorter timescales.

Shortcuts-to-adiabaticity can also be achieved via incoherent dynamics \chng{for state manipulation~\cite{Vacanti:2014} and can be a useful tool to speed up thermalisation and equilibration~\cite{ChenuPRR}.} For example, an ideal pure dephasing channel enhances adiabaticity by performing an effective projection on the instantaneous basis~\cite{avron2011landau}. 
In order to realise the required pure dephasing, one can engineer the coupling of the system to an auxiliary system or ``meter'' \cite{raphael}, in the spirit of a quantum non-demolition (QND) measurement~\cite{haroche2006exploring}. Since the interaction in a QND measurement commutes with the system Hamiltonian at all times, it affects only system coherences, not populations. Evidence for enhancing adiabaticity in the Landau-Zener model has been demonstrated numerically~\cite{raphael}, where for a time-dependent system Hamiltonian, as in annealing protocols, the QND coupling necessarily also becomes time-dependent. 

Here we propose a protocol inspired by the QND Hamiltonian that achieves enhanced adiabaticity from a mechanism that, remarkably, is entirely coherent.
We refer to this mechanism as energy rescaling since, roughly speaking, an effective widening of energy gaps in the system Hamiltonian is the key and it is achieved due to the energy added by coupling to the meter. The mechanism arises from the necessary condition that the interaction between system and meter creates dephasing, namely to commute with the system Hamiltonian. 
We show analytically and confirm numerically that:
\begin{enumerate}\setlength{\itemsep}{0pt}
    \item dephasing-generating protocols change the spectral range of the system;
    \item the maximal speedup of a dephasing-generating protocol is achieved when system and meter are at all times in a separable state;
    \item the achievable speedup is linearly proportional to the interaction strength between system and meter, in both the adiabatic and non-adiabatic regimes.
\end{enumerate}
Given the fact that our protocol does not necessarily generate correlations between system and meter, strictly speaking the labelling ``dephasing'' is improper. Nevertheless, a typical QND protocol reduces to our model when the energy of the QND coupling exceeds the meter eigenenergy. We therefore refer to our protocols as ``QND-like''.
In the limit of the QND coupling being much larger than the meter eigenenergy, our results show that the QND coupling effectively induces a pure energy rescaling. The corresponding protocol provides a speedup in terms of an overall factor scaling the time of the adiabatic transition.
Our findings highlight some aspects of the counter-intuitive nature of engineered dephasing. We also provide analytical guarantees for the rate of speedup that our protocols can achieve in quantum annealing scenarios. 

The remainder of paper is structured as follows: We shortly review quantum annealing and introduce our protocol in Sec.~\ref{sec:QA+QND}. In Sec.~\ref{sec:mechanism}, we show that our protocol can induce pure dephasing, but also 
induces energy rescaling, and demonstrate that the latter effect is the cause of enhanced adiabaticity. 
We derive an upper bound on the speedup due to the enhanced adiabaticity in Sec.~\ref{sec:mechanism}. 
In Sec.~\ref{sec:speedup} we establish that our protocol leads to a linear speed up regardless of whether we are in the adiabatic or non-adiabatic regime.
In Section~\ref{sec:protocol}, we discuss requirements for realizing the QND-like 
protocol for quantum annealing applications, introduce a simplified protocol 
with fewer types of physical interactions, and determine the performance with a numerical benchmark. Finally, we conclude our results in Sec.~\ref{sec:concl}.

\section{Quantum annealing, dephasing, and quantum non-demolition}
\label{sec:QA+QND}

We begin by introducing the general class of problems that will be the focus of this work and briefly review the QND-type proposal of Ref.~\cite{raphael}. In the context of combinatorial optimization, QA is usually used to find solutions to a problem called Quadratic Unconstrained Binary Optimization (QUBO), relevant in finance, logistics, machine learning, and other fields, and is an NP-hard problem~\cite{glover2019quantum}. Furthermore, there are ways to express some other NP-hard problems, e.g. vertex cover, MAXCUT, and graph colouring, as QUBO problems~\cite{glover2019quantum} and it is possible to transform optimization problems into this form in an automated way~\cite{AutomaticQUBO}.
The problem is the following: given $N$ binary variables ${x_i}: x_i \in \{0,1\}$, and an $N \times N$ real matrix $Q$, find the $\{x_i\}$ corresponding to the extremum of
\begin{equation}
    y(x) = \sum_{i=1}^{N} \sum_{j=1}^{N} Q_{ij} x_i x_j.
\end{equation}
With a simple substitution of rescaled variables, $z_i = 2x_i - 1 \in \{-1,1\}$, one can quickly see that the solution to a QUBO problem corresponds to the ground state of an Ising Hamiltonian~\cite{glover2019quantum}, 
\begin{equation}
\label{eqn:Ising}
    H_I = \sum_{ij} J_{ij} \sigma_z^i \sigma_z^j + \sum_i h_i \sigma_z^i\,,
\end{equation}
where the coefficients $J_{ij}=Q_{ij}/4$ and $h_i=\sum_j{Q_{ij}/2}$ are found from the problem matrix, $Q$. Finding the ground state is precisely the goal of QA, and it is done by first preparing the system in the ground state of a simpler transverse field Hamiltonian, $H_T = \sum_i \sigma_x^i$, e.g. by cooling, which is subsequently evolved under the time dependent Hamiltonian, 
\begin{equation}\label{eq:H_QA}
    H(t) = (1-f(t))H_T + f(t)H_I\,,
\end{equation}
where $f(t)$ is some switching function satisfying $f(0) \!=\! 0$ and $f(t_{final}) \!=\! 1$. Then, for a slow enough evolution, the system will stay in its ground state by the adiabatic theorem, and thus measuring it at the end of the protocol will yield the solution to the QUBO problem. 

To prevent diabatic transitions, i.e., transitions out of the instantaneous ground state, dephasing can be added to the system dynamics. Dephasing refers to the decay of the off-diagonal elements of the density matrix, i.e. the coherences. This definition is, of course, basis-dependent. In the following (and as is common), we shall refer to dephasing with respect to the instantaneous eigenbasis of the system, i.e., the time-dependent basis that diagonalises the system Hamiltonian at every instant in time.

Intuitively, one can think of the adibaticity enhancement caused by dephasing by noting that diabatic transitions create superpositions between the ground state and some excited state(s). The superposition results in coherences in the density matrix. Since dephasing causes these coherences to decay, it can also suppress them from forming, thus maintaining the system in its ground state. Another way to think about the effect of dephasing is through the analogy to the quantum Zeno effect~\cite{itano1990quantum}. It refers to frequent repeated measurements ``freezing'' the system in an eigenstate of a possibly time dependent observable; in our case, the energy of the system. Dephasing is the time-continuous version of a quantum measurement and is thus capable of freezing a system in its eigenstate for sufficiently strong dephasing rates.

Engineering such dissipation is highly non-trivial as it must operate in a time-dependent basis that coincides, at all times, with the energy eigenbasis of the system. One way to achieve this to couple the system to an auxiliary system, a quantum meter, through an interaction that commutes with the system Hamiltonian at all times. No energy can be exchanged between system and meter, and therefore only coherences in the system eigenbasis are impacted, while populations remain unaffected. This can be implemented with a QND protocol~\cite{raphael}, where the system-meter interaction replicates the system Hamiltonian, $H_{int}(t) = H_S(t) \otimes X_M$ for some operator $X_M$ in the meter Hilbert space, similar to a QND Hamiltonian~\cite{haroche2006exploring}. The total Hamiltonian of system and meter is then
\begin{equation}
\label{eq:mainHamiltonian}
    H(t) = H_S(t) \otimes \mathds{1} + H_S(t) \otimes X_M + \mathds{1} \otimes H_M\,,
\end{equation}
where $H_M$ is the inherent Hamiltonian of the meter.

The QND protocol and its performance were analysed numerically~\cite{raphael} and analytically \cite{King:2024} for the particular case where the meter is a harmonic oscillator that itself is coupled to a thermal reservoir and the system is a qubit. It was shown that this kind of protocol does not provide a speedup in terms of computational complexity with respect to coherent, adiabatic transitions \cite{King:2024}. Numerical simulations, nevertheless, indicated a  speedup of the dynamics by increasing the strength of the coupling with the meter \cite{King:2024}. The asymptotic acceleration scaled linearly with the coupling, indicating a speedup in terms of an overall, constant factor scaling the time. 
Here, we assess the speedup by considering a solvable model, where the QND coupling Hamiltonian commutes with the instantaneous Hamiltonian. This model reproduces the dynamics of Eq.\ \eqref{eq:mainHamiltonian} for large QND couplings, when $H(t)\approx H_S(t)\otimes X_M$, and its predictions shed light onto this asymptotic limit.

\section{Dephasing and energy rescaling mechanisms}
\label{sec:mechanism}
We now consider a QND-like protocol that can induce an effective dephasing on the system, and show that the speedup it achieves is an overall global factor rescaling the time.

\subsection{Dephasing}
\label{subsec:dephasing_mechanism}

We consider a simple example using a qubit meter and in what follows show that the our protocol results in dephasing for the system. Suppose the meter is  initialised in $\ket{+} = \frac{1}{\sqrt{2}} (\ket{0}+\ket{1})$, where $\sigma_z\ket{0}=\ket{0}$ and $\sigma_z\ket{1}=-\ket{1}$ and we set $X_M = x_0 \sigma_z$, $H_M = 0$ but keep the $H_S(t)$ general for the purposes of the derivation. Here, $x_0$ is a real number scaling the strength of the QND coupling. The total Hamiltonian is then
\begin{eqnarray}
\label{eqn:H}
  H(t) & = &H_S(t) \otimes \mathds{1} + x_0 H_S(t) \otimes \sigma_z\,, \\ & =& (1+x_0)H_S(t) \otimes \ket{0}\bra{0} + (1-x_0) H_S(t) \otimes \ket{1}\bra{1}\,.                                                                      \nonumber
\end{eqnarray}
The time evolution operator for this Hamiltonian, $U(t)$, is
\begin{equation}
    U(t) = U^{[1+x_0]}_{QND}(t) \otimes \ket{0}\bra{0} + U^{[1-x_0]}_{QND}(t) \otimes \ket{1}\bra{1}\,,
\end{equation}
where $U^{[1 \pm x_0]}_{QND}(t)$
denotes the time evolution operator for a rescaled system Hamiltonian $(1 \pm x_0) H_S(t)$ (with the rescaling coming from, as far as the system is concerned, a QND-type coupling, hence the subscript), i.e., 
\begin{equation}
\label{eqn:U_rescaled}
    \frac{d}{dt}U^{[x]}_{QND}(t) = ixH_S(t)U^{[x]}_{QND}(t),
\end{equation}
where $x=1\pm x_0$.
Suppose that system and meter are initially in the separable state $\rho_S(0) \otimes \ket{+}\bra{+}$. We now perform a Kraus decomposition by tracing out the meter. Since the meter density matrix has only one non-zero eigenstate $\ket{+}$, there are only two non-zero Kraus operators,
\begin{subequations}\label{eq:Kraus}
\begin{eqnarray}
    K_{++}(t) &=& \frac{1}{\sqrt{2}}\left(U^{[1+x_0]}_{QND}(t) + U^{[1-x_0]}_{QND}(t)\right)\,, \\
    K_{-+}(t) &=& \frac{1}{\sqrt{2}}\left(U^{[1+x_0]}_{QND}(t) - U^{[1-x_0]}_{QND}(t)\right)\,.
\end{eqnarray}
\end{subequations}    
The system density matrix is then given by
\begin{equation}
    \rho_S(t) = K_{++}(t) \rho_S(0) K_{++}^\dag(t) + K_{-+}(t) \rho_S(0) K_{-+}^\dag(t)\,.
  \end{equation}
Substituting with Eqs.~\eqref{eq:Kraus} and simplifying, we obtain
\begin{eqnarray}
\label{eqn:rhoU}
        \rho_S(t) &= & \frac{1}{2} U^{[1+x_0]}_{QND}(t) \rho_S(0) \left(U^{[1+x_0]}_{QND}(t)\right)^\dag \\ \nonumber  &&+  \frac{1}{2} U^{[1-x_0]}_{QND}(t) \rho_S(0) \left(U^{[1-x_0]}_{QND}(t)\right)^\dag.
\end{eqnarray}
Equation~\eqref{eqn:rhoU} implies that half of the initial state evolves under $(1+x_0)H_S(t)$, while the other half evolves under $(1-x_0)H_S(t)$, and then the two results are superposed:
\begin{equation}
\label{eqn:rho}
    \frac{d}{dt}\rho_S(t) = \frac{1}{2} \frac{d}{dt} \rho^{[1+x_0]}_S(t) + \frac{1}{2} \frac{d}{dt} \rho^{[1-x_0]}_S(t)\,,
\end{equation}
where $\rho^{[1 \pm x_0]}_S(t)$ is the density matrix of a system evolving under $(1 \pm x_0)H_S(t)$. 
In Appendix~\ref{Appendix1} we show that we can rewrite Eq.~\eqref{eqn:rho} in the instantaneous eigenbasis of $H_S(t)$ which results in the expression
\begin{widetext}
\begin{eqnarray}\label{eq:corr1}
    \frac{d}{dt} \left(\rho_S(t)\right)_{mn} &=& -i\left(\left[H_S(t), \rho_S(t)\right]\right)_{mn} 
    + \frac{i x_0}{2} \left(E_m(t) - E_n(t)\right)\left((\rho^{[1 + x_0]}_S(t))_{mn} - (\rho^{[1 - x_0]}_S(t))_{mn}\right)\,.
\end{eqnarray}
\end{widetext}
From Eq.~\eqref{eq:corr1}, it is evident that the interaction with the meter yields a correction in the density matrix evolution that, to first order, only affects the off-diagonal elements in the instantaneous eigenbasis due to the $(E_m - E_n)$ term. This correction could either increase or decrease the off-diagonal elements (i.e. the coherences), but since the coherence of a pure state is already maximal, only the latter effect is possible. This means that coherences are suppressed i.e. dephasing is induced. Furthermore, since we have shown this through an exact Kraus decomposition,  the result also holds for strong interactions, fast (non-adiabatic) evolutions of the system, and in the presence of non-Markovian effects.

\begin{figure}[tbp]
    (a)\\[-3ex]
    \includegraphics[width=0.9\columnwidth]{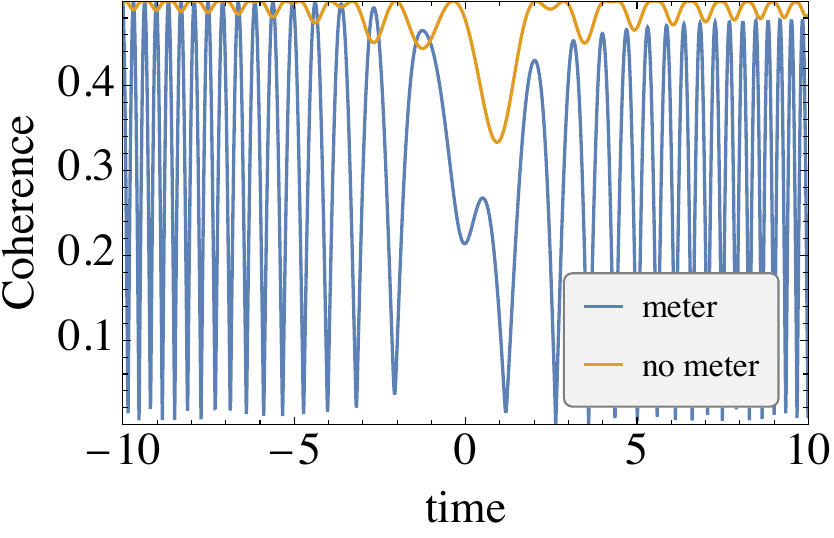}\\
     (b)\\[-3ex]
    \includegraphics[width=0.9\columnwidth]{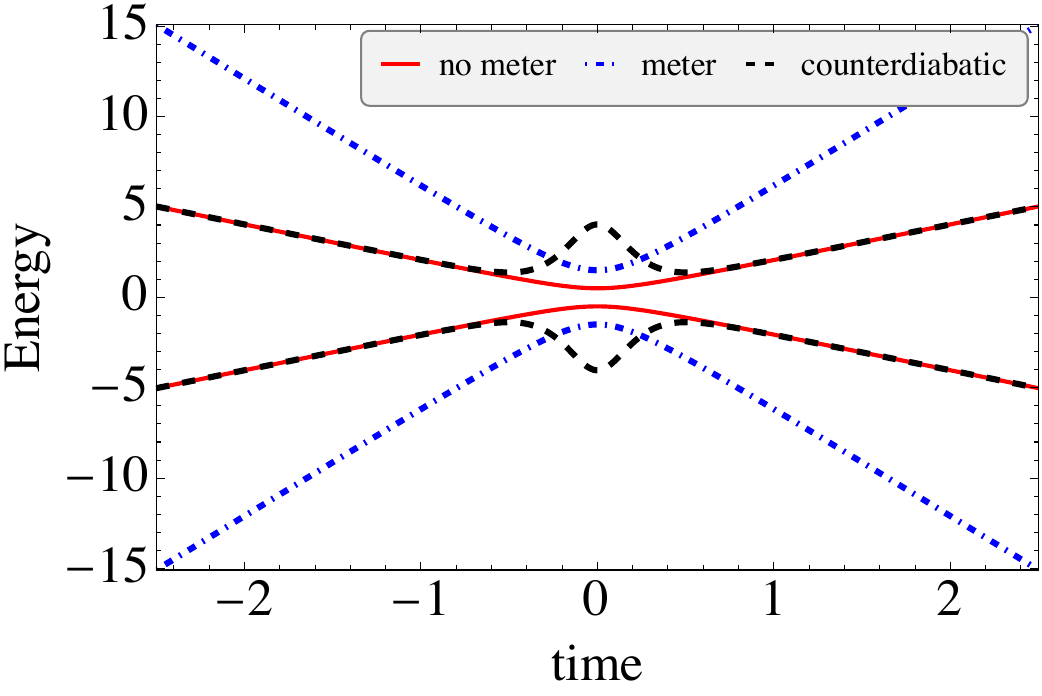}
    \caption{(a) Magnitude of coherence as a function of time, with a quantum meter (QND-like protocol) and without it (coherent protocol). The system, evolving under the Landau-Zener Hamiltonian, is initialised in a uniform superposition of instantaneous ground and first excited state, and the magnitude of the coherence between them is plotted against time; the periodic suppression of coherence indicates dephasing created by our protocol. (b) Energy spectra for the bare LZ model (no meter, red) and QND-like protocol (with meter, dot-dashed blue) which shows the energy rescaling induced by coupling to the meter. For comparison we also show the behavior of the energy spectrum when counterdiabatic control~\cite{berry_cd, AbahNJP} is employed (black, dashed). We fix $T=5$, $x_0=2$, and $g=1$.}
    \label{fig:coherences}
\end{figure}

The above illustrates that the QND coupling induces pure dephasing in the instantaneous basis of the meter. In what follows we determine the strength and time-dependence of the dephasing rate. It is therefore informative to consider an example setting, allowing to determine the strength and time-dependence of dephasing numerically. We consider the Landau-Zener (LZ) Hamiltonian for a single qubit as in Ref.~\cite{raphael},
\begin{equation}
\label{eqn:LZ}
    H_S(t)=H_{LZ}(t)=\frac{vt}{2} \sigma_z + \frac{g}{2} \sigma_x\,,
\end{equation}
where the meter is coupled to the system via $H_{int} = H_S(t) \otimes \sigma_z$ and we set $H_M = 0$ here for simplicity. We initialise system and meter in $\ket{+} \otimes \ket{+}$, such that the system starts in a uniform superposition of ground and excited states. We linearly ramp the system from $t = -10/v$ to $t=+10/v$ with $v = g = 1$, trace out the meter and plot, in Fig.~\ref{fig:coherences}(a), the magnitude of the coherence of the density matrix element, $\rho_{01}(t)$, as a function of time. Comparing it to the case where the meter is absent, $H_{int}\!=\! 0$, we observe a speed up of the oscillation of the coherence in the instantaneous basis, as well as an increase of the oscillation amplitude. Averaging over time gives rise to a net reduction of the coherence with respect to the dynamics in the absence of the meter, indicating that the meter induces a sort of dephasing on the qubit's dynamics.

\subsection{Energy rescaling}

A natural generalization is to consider any interaction $Y_S(t) \otimes X_M$ that commutes with the system Hamiltonian at all times, $[Y_S(t), H_S(t)]=0$. In fact, every interaction that induces pure dephasing in the system eigenbasis is of this form. More specifically, any interaction that does not satisfy the commutation relation also induces relaxation in the system eigenbasis, for the following reason: Given the eigenbasis of $H_S(t)$, $\{\ket{\psi_i(t)}\}$, the interaction can be written as $$\left(\sum_{ij} Y_{ij}(t) \ket{\psi_i(t)} \bra{\psi_j(t)}\right) \otimes X_M.$$ 
If $[Y_S(t), H_S(t)] \neq 0$, then there exists a matrix element $Y_{mn}(t) \neq 0$ for $m \neq n$. However, this implies that that the $m^\text{th}$ and $n^\text{th}$ eigenstates are coupled by $Y_S(t)$, and therefore after tracing out the meter one would in general obtain relaxation between these eigenstates. 

The commuting property leads to another relevant mechanism. Since the system Hamiltonian commutes with the interaction, they can be simultaneously diagonalised. Therefore, there exists a particular eigenbasis of the system $\{\ket{\psi_i(t)}\}$, where one can write the total Hamiltonian as
\begin{equation}
  H(t)=  \sum_{i}\ket{\psi_i(t)}\bra{\psi_i(t)} \otimes (E_i(t) \mathds{1} +y_i(t) X_M)\,,
\end{equation}
where $y_i(t)$ is the eigenvalue of $Y_S(t)$ corresponding to $\ket{\psi_i(t)}$ and for simplicity we have assumed $H_M = 0$ and remark that this condition will be relaxed in Sec~\ref{sec:speedup}. 
Let $\{\ket{m_j}\}$ be the eigenbasis of $X_M$ with corresponding eigenvalues $\{m_j\}$. We can write the total Hamiltonian in a doubly block-diagonal form,
\begin{equation}
    H(t) = \sum_{i,j} (E_i(t) + y_i(t) m_j)\ket{\psi_i(t)}\bra{\psi_i(t)} \otimes \ket{m_j}\bra{m_j}\,.
\end{equation}
The time evolution operator can then be written as 
\begin{equation}
\label{eqn:U}
    U(t) = \sum_j U_R^{[j]}(t) \otimes \ket{m_j}\bra{m_j}\,,    
\end{equation}
where $U_R^{[j]}(t)$ is the \textit{rescaled time evolution operator}, defined as the time evolution operator corresponding to the rescaled Hamiltonian, $H_R^{[j]}(t)$,
\begin{equation}
    \frac{d}{dt}U_R^{[j]}(t) = i H_R^{[j]}(t) U_R^{[j]}(t)\,,
  \end{equation}
  with 
\begin{equation}
    H_R^{[j]}(t) = \sum_{i} (E_i(t) + y_i(t) m_j)\ket{\psi_i(t)}\bra{\psi_i(t)}.
\end{equation}
This is simply a more general version of the rescaled Hamiltonian introduced in Sec.~\ref{sec:QA+QND} which is 
recovered by setting $Y_S(t)\!=\!x_0 H_S(t)$. If system and meter are initialised in a product state, $\rho_S(0) \otimes \rho_M(0)$, using Eq.~\eqref{eqn:U} we can perform a Kraus decomposition and obtain the evolution of the system alone, $\rho_S(t) = \sum_j K_j(t) \rho_S(0) K_j^\dag(t)$, where
\begin{equation}
    K_j(t) = \sqrt{\bra{m_j}\rho_M(0)\ket{m_j}} U_R^{[j]}(t).
\end{equation}
The evolution of the reduced density matrix for the system is now readily understood: $\rho_S(0)$ gets split up into $dim(X_M)$ parts, corresponding to every eigenstate of $X_M$, weighted by its overlap with the initial state, $\bra{m_j}\rho_M(0)\ket{m_j}$. For every eigenstate, each part is evolved under the corresponding rescaled Hamiltonian $H_R^{[j]}(t)$, and finally, the parts are added back up to yield $\rho_S(t)$. Having written explicitly the time evolution, the effect of the interaction becomes clear; it shifts the energies of the eigenstates of $H_S(t)$. As a result, this affects the rate of diabatic transitions from the ground state and therefore the fidelity of the annealing protocol. If the interaction increases the gap between the ground state and another eigenstate transitions are suppressed, while if the interaction reduces the gap they are enhanced.

Through the perspective of shifting energies of instantaneous eigenstates, the utility of our QND-like protocol can be more thoroughly understood. It is a specific case of protocols with $Y_S(t) \!=\! H_S(t)$; then, $y_i(t) \!=\! E_i(t)$ and the energy gaps are \textit{rescaled}. In other words, if the meter is initialised in an eigenstate of $X_M$ with a positive eigenvalue $m_j \!>\! 0$, the energy gaps between every pair of eigenstates of $H_S(t)$ are enlarged by a factor of $1+m_j$. The larger this factor is, the more diabatic transitions out of the ground state are suppressed. One could try to engineer an interaction designed specifically to enlarge the gap between the instantaneous ground and first excited states, for instance, $Y_S(t) = \Delta_E(t) (\ket{\psi_1(t)} \bra{\psi_1(t)} - \ket{\psi_0(t)} \bra{\psi_0(t)})$. 
Indeed, this is essentially how counterdiabatic protocols achieve their speedup~\cite{berry_cd}. For comparison in Fig.~\ref{fig:coherences}(b) we show the energy spectra for the bare LZ (red), the QND-like approach proposed in this work (dot-dashed, blue) and the case of control achieved via a counterdiabatic Hamiltonian (dashed, black)~\cite{berry_cd,AbahNJP} \chng{for the same linear protocol as used in panel (a)}. It should be immediately apparent that our approach maintains the qualitative spectral features of the bare model. While clearly both our QND-like and counterdiabatic approaches achieve a finite time adiabatic dynamics by suitably adjusting the spectral properties of the model, the manner in which this is performed is markedly different. An important distinction between the two settings is that the spectral adjustments necessary for counterdiabatic driving require knowing the exact eigenstates and eigenvalues of the system Hamiltonian.
However, in an annealing scenario, the eigenstates and eigenvalues of the system Hamiltonian are not known {\it a priori}, as otherwise the solution to the optimization problem at hand would already be available. Clearly, this limits the applicability of certain STA approaches to QA type problems (we will discuss this in more detail in Sec.~\ref{STAcomparisons}). Therefore, in what follows we focus on the properties of the QND-like protocol, as it is the simplest dephasing protocol that is both useful, insofar as it is guaranteed to increase the energy gaps, and computationally feasible as it does not require explicit prior knowledge of the eigenstates of $H_S(t)$.

With energy rescaling in mind, we can easily identify the optimal initial meter state $\rho_M(0)$ to maximise speedup of the our protocol---one should pick an eigenstate of $X_M$ with maximal eigenvalue since this will maximise the energy rescaling factor $(1+m_j)$ and, thus, suppresses diabatic transitions by the largest extent. For $H_M\!=\!0$ this choice of $\rho_M(0)$ leads to completely unitary dynamics, as there would be only one non-zero Kraus operator. Remarkably, this implies that the optimal regime of these protocols corresponds to one without any dephasing whatsoever.

\section{Quantifying the speedup}
\label{sec:speedup}

Having gained insight into the speedup provided by our QND-like protocol, we now seek to quantify it. Following~\cite{roland2002quantum}, we ensure adiabaticity by replacing $H_{tot}(t)$ with $H_{tot}(s(t))$, where $s(t)$ is defined as the function slowing down the evolution of $H_{tot}(t)$ in such a way that it is locally adiabatic at all times. Quantitatively, this means that the protocol will stay adiabatic up to an error of $\epsilon^2$, provided
\begin{equation}
\label{eqn:adiabaticity}
    \frac{ds}{dt} \frac{|M|}{g^2(t)} \leq \epsilon.
\end{equation}
Here, $M$ is the matrix element of $\dot{H}_{tot}(s(t)) = \dot{H}_S(s(t)) \otimes (\mathds{1} + X_M)$ connecting the instantaneous ground state and another instantaneous eigenstate, which is the lowest lying eigenstate for which $M$ is non-zero, and $g(t)$ is the energy gap between those two states. Consider the case when $X_M$ and $H_M$ commute (which evidently holds when $H_M = 0$) and the meter is initialised in $\ket{m_i}$, an eigenstate of $X_M$ and therefore also $H_M$. If the system is adiabatic, the state will remain (up to a phase factor) equal to $\ket{\psi_0(t)} \otimes \ket{m_i}$. 
Since $\dot{H}_{tot}(s) \!=\! \dot{H}_s(s) \otimes (\mathds{1} + X_M)$ is diagonal in the eigenbasis of $X_M$ and $H_M$, it will only connect states with the same meter eigenstate. Therefore $M$ connects $\ket{\psi_0(t)}\otimes\ket{m_i}$ and $\ket{\psi_1(t)} \otimes \ket{m_i}$:
\begin{equation}
\begin{split}
        M & = [\bra{\psi_0(t)} \otimes \bra{m_i}] \dot{H}_{tot}(s(t)) [\ket{\psi_1(t)} \otimes \ket{m_i}] \\ & =  \bra{\psi_0(t)} \dot{H}_{s}(s(t)) \ket{\psi_1(t)} \bra{m_i} (\mathds{1} + X_M) \ket{m_i} \\ & = (1+m_i)\bra{\psi_0(t)} \dot{H}_{s}(s(t)) \ket{\psi_1(t)}\,,
\end{split}
\end{equation}
where $X_M\ket{m_i} = m_i\ket{m_i}$. Hence, $g(t)$ is simply $(1+m_i)(E_1(t) - E_0(t))$, which cancels the $(1+m_i)$ term in the numerator of Eq.~\eqref{eqn:adiabaticity} and the adiabaticity factor $\frac{|M|}{g^2(t)}$ is consequently reduced by $(1+m_i)$. This implies that the annealing time is reduced by the same factor when compared to the coherent case without coupling to the meter. Thus, under the assumption $[X_M, H_M] = 0$, a fully adiabatic annealing schedule can be implemented with a speedup of at most $1+m_{max}$, where $m_{max}$ is the maximal eigenvalue of $X_M$.

\begin{figure}[t]
    (a) \\[-3ex]
    \includegraphics[width=0.8\columnwidth]{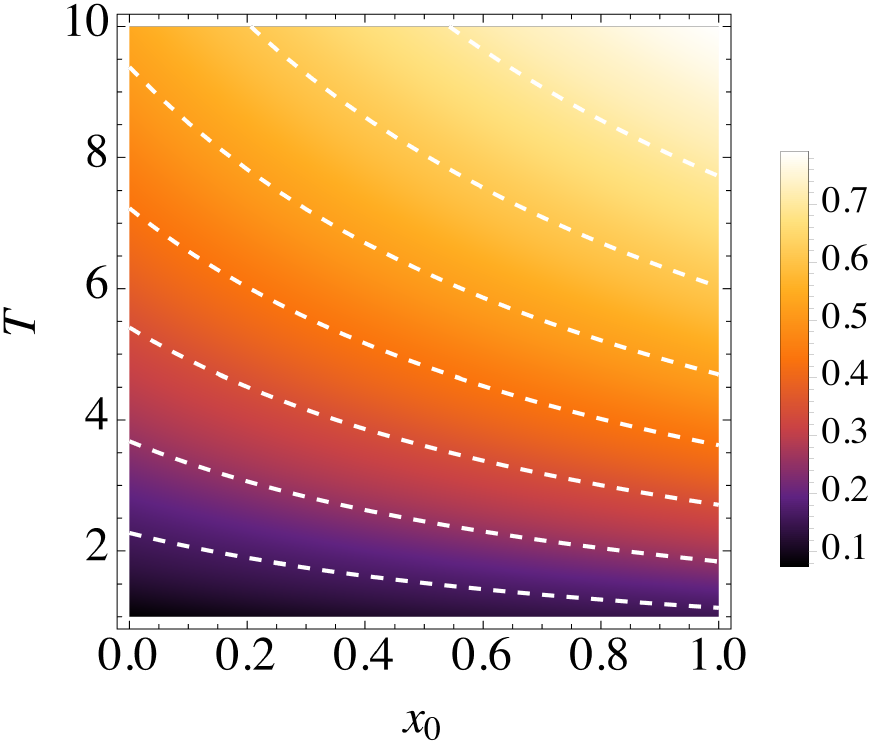}\\
    (b)\\[-3ex]
    \includegraphics[width=0.8\columnwidth]{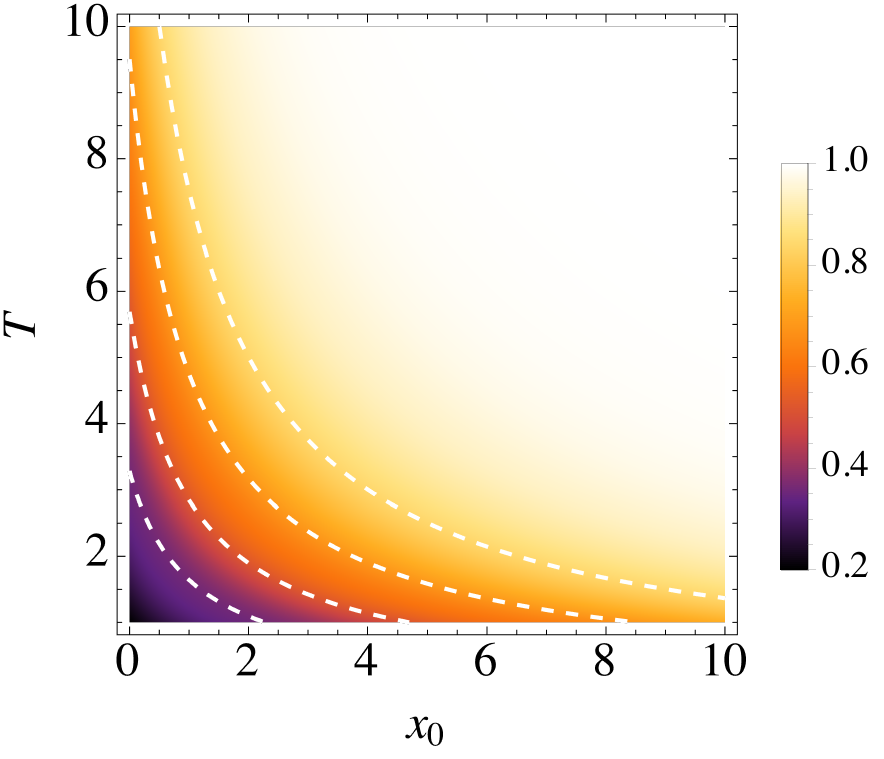}
    \caption{Fidelity of our QND-like protocol for two example system Hamiltonians $H_S(t)$, calculated for various annealing durations, $T$, and system-meter interaction strengths, $x_0$, (with $X_M = x_0 \sigma_z$, $H_M = 0$). Dashed lines show examples of $T(1+x_0) = \text{const}$. The lines perfectly follow regions of constant fidelity even when the fidelity is low, i.e., in the non-adiabatic regime, indicating the same scaling of the speedup with system-meter interaction as proven for the adiabatic regime. (a) system is a qubit evolving under Landau-Zener Hamiltonian (Eq. ~\eqref{eqn:LZ} with $g=1$). (b) system is an array of $N=3$ qubits evolving under Eq. ~\eqref{eq:H_QA} with $H_f$ set to be the Ising Hamiltonian (Eq.~\eqref{eqn:Ising}).}
    \label{fig:adiabaticity:commut}
\end{figure}
We now numerically test the validity of this result outside the adiabatic regime. To this end, we pick two example systems: a qubit evolving under the Landau-Zener Hamiltonian Eq.~\eqref{eqn:LZ} and a qubit array with $N\!=\!3$ undergoing an annealing schedule Eq.~\eqref{eq:H_QA} to find the ground state of an Ising Hamiltonian Eq.~\eqref{eqn:Ising} with uniformly random parameters $J_{ij}\!\in\!\left[0,1\right]$ and $h_i\!\in\!\left[0,1\right]$. In both cases, the meter is a qubit with inherent Hamiltonian $H_M \!=\! \omega \sigma_x$ and is coupled to the system via $H_{int} \!=\! x_0 H_S(t) \otimes \sigma_z$. 
We initialise the meter in $\ket{0}$, which we remind is the eigenstate of $\sigma_z$ corresponding to eigenvalue 1,
such that it induces positive energy rescaling. The system is in its ground state, and propagates for an annealing duration $T$. The initial and final Hamiltonians should be independent of $T$. For the single-qubit case, this is achieved by linearly ramping the system from $t\! =\! -10/v$ to $t\!=\! +10/v$, such that $T \!=\! 20/v$; while for the qubit array case this is guaranteed automatically from the form of Eq.~\eqref{eq:H_QA} \chng{for the considered linear protocol $f(t)=t/T$}. 
We determine the success of the protocol by calculating the fidelity between the actual (target) ground state of the system and the state at the end of the protocol, $\mathcal{F}=\vert \bra{\psi_0(T)} \psi(T) \rangle \vert^2$.

Figure~\ref{fig:adiabaticity:commut} displays the fidelity as a function of protocol duration $T$ and system-meter interaction strength $x_0$, where we fix $\omega=0$ and therefore $H_M=0$, such that $[X_M, H_M] = 0$ for both considered systems. We observe that for any value of $T$, the fidelity for $x_0=0$ is the same as for $T' \!=\! T(1+x_0)$. In other words, the fidelity is constant along the line $T(1+x_0)$, examples of which are delineated by the white dashed lines in Fig.~\ref{fig:adiabaticity:commut}. This is also true when the value of the fidelity is appreciably lower than one, i.e. in the non-adiabatic regime, thus demonstrating that the same speedup is achieved even when going faster than adiabatically.

\begin{figure}[tbp]
    \includegraphics[width=0.45\textwidth]{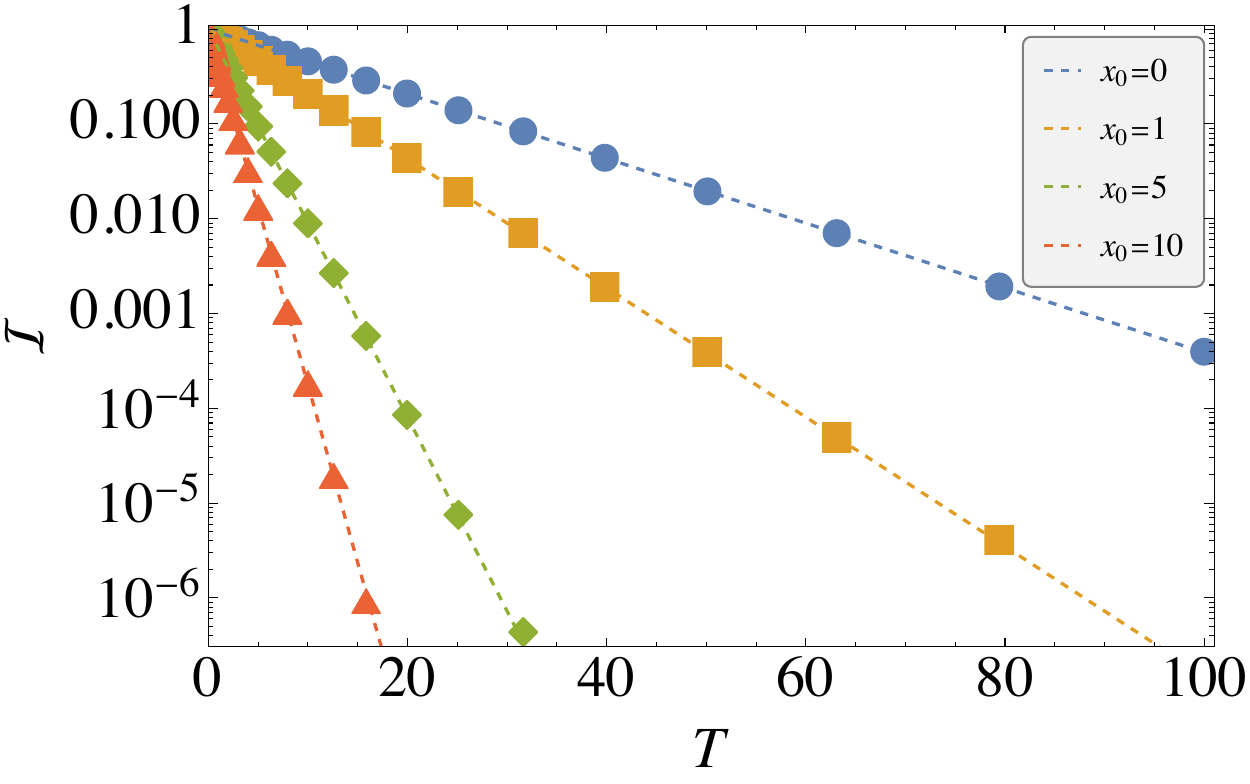}
    \caption{Final fidelity, (we show $\mathcal{I}=1-\mathcal{F}$), between final state from the QND-like protocol with target ground state of $H_S(t_f)$ for various values of system-meter coupling strengths $x_0$ (data points). The lines show the predicted LZ fidelity~\cite{Landau1932, Zener1932, GarrawayPRA} demonstrating perfect agreement when the total protocol duration is re-scaled from $T\to T(1+x_0)$ which is equivalent to an energy rescaling for the LZ model.}
    \label{fig:LZ_infidelities}
\end{figure}

For the two-level case, we can readily demonstrate the validity of the above claim in the non-adiabatic regime by considering the well known LZ formula~\cite{Landau1932, Zener1932, GarrawayPRA},
\begin{equation}
\mathcal{I}=1-\mathcal{F}=\text{exp}\left[-\pi \frac{(g/2)^2}{v/2} \right]\,
\end{equation}
which predicts the fidelity of the evolved state with the adiabatic state for a finite time ramp. Due to the simplicity of the LZ model, the energy rescaling can equivalently be thought of as a rescaling of the effective timescale of the drive. In Fig.~\ref{fig:LZ_infidelities} we consider the same protocol for the LZ as previously, i.e. the system is evolved from $t\! =\! -10/v$ to $t\!=\! +10/v$ with $v=20/T$ and show that the QND-like protocol gives precisely the expected LZ scaling of the final state fidelity when $T$ is transformed according to $T\to T(1+x_0)$, thus showing that the linear enhancement in speed up is achieved regardless of how fast the system is being driven.

\begin{figure}[t]
    (a) \\[-3ex]
    \includegraphics[width=0.87\columnwidth]{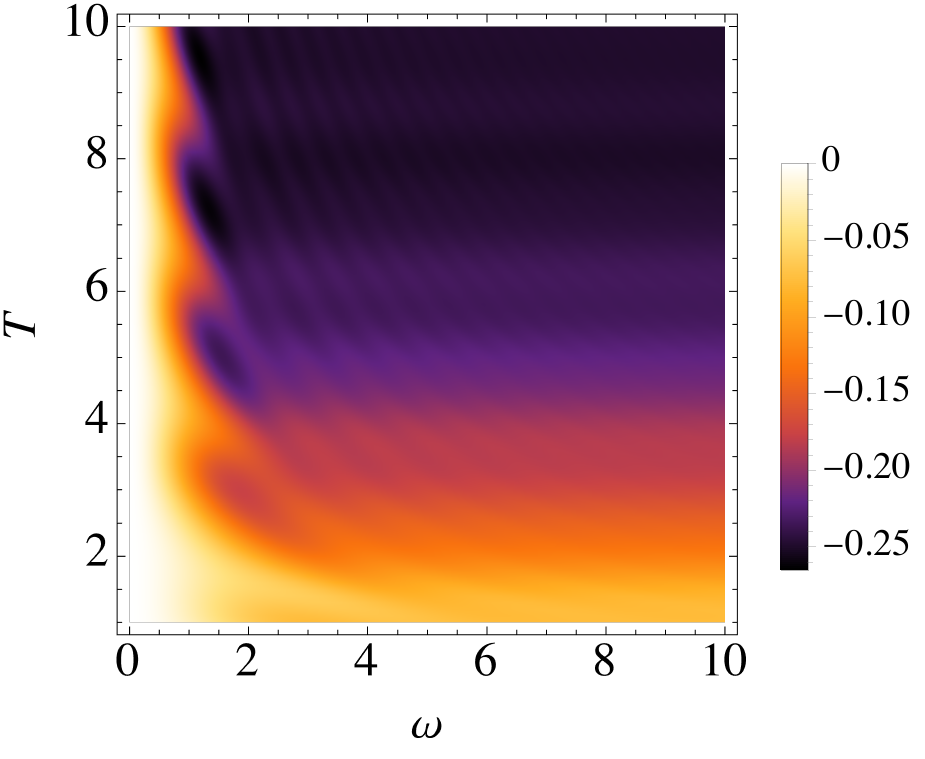}\\
    (b)\\[-3ex]
    \includegraphics[width=0.87\columnwidth]{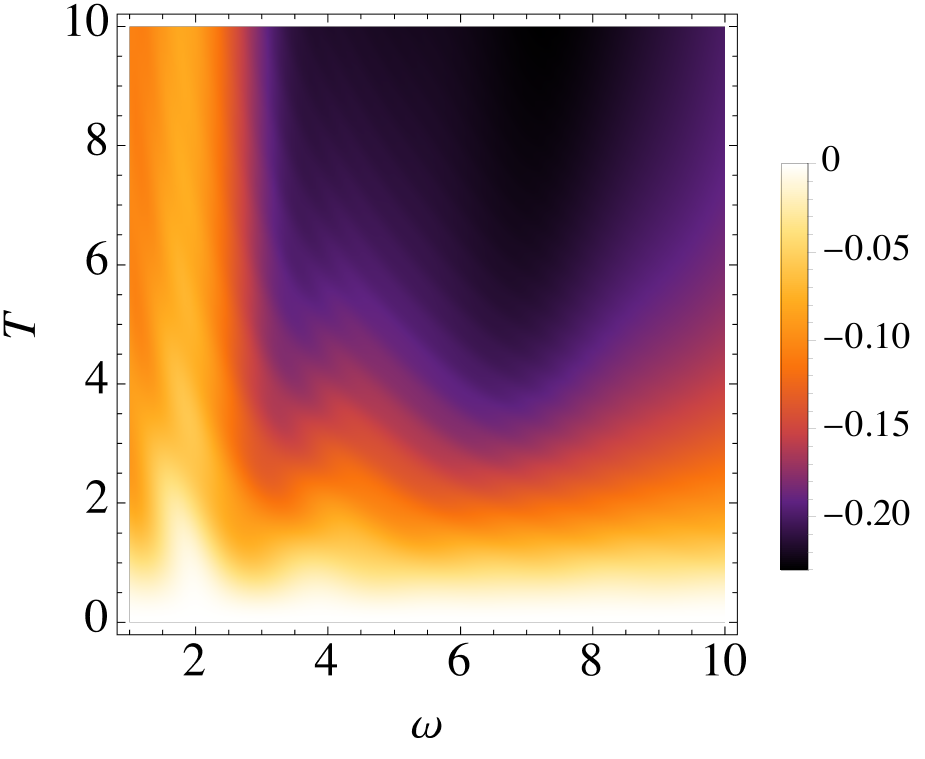}
    \caption{Fidelity difference between the QND-like protocols with $H_M = 0$ and $H_M = \omega \sigma_x$, as a function of protocol duration $T$ and strength $\omega$ of the meter Hamiltonian. The larger the value of $\omega$, the higher the magnitude of $[X_M, H_M]$. The fidelity difference is always negative, indicating the speedup of the non-commuting case to be upper bounded by the commuting case. (a) System is a qubit evolving under Landau-Zener Hamiltonian (Eq.~\eqref{eqn:LZ} with $g=1$). (b) system is an array of $N=3$ qubits evolving under Eq.~\eqref{eq:H_QA} with $H_f$ set to be the Ising Hamiltonian (Eq.~\eqref{eqn:Ising}). In both panels we fix the system meter coupling such that $x_0=1$.}
    \label{fig:adiabaticity:noncommut}
\end{figure}

Having provided evidence for the validity of our result beyond the adiabatic limit when $X_M$ and $H_M$ commute, we now conjecture that $1+m_{max}$ is an upper bound for the achievable speedup also in the non-commuting case, $[X_M, H_M] \neq 0$. The intuition follows since if the meter is initialised in its optimal state, i.e. the eigenstate of $X_M$ with maximal eigenvalue, then the presence of a meter Hamiltonian $H_M$ which does not commute with $X_M$ will kick the meter out of this state, leading to a smaller energy rescaling effect. 
To evaluate the effect of the non-commuting meter Hamiltonian, we use the same two example systems and, fixing $x_0\!=\!1$,  scan over different values of $\omega$ and $T$. For each $T$, we calculate the difference in fidelity between a protocol with a finite $\omega$ and the protocol with $\omega = 0$. In other words, we plot the quantity $q(T, \omega) - q(T, 0)$, where $q$ is the protocol fidelity. The result is shown in Fig.~\ref{fig:adiabaticity:noncommut}. We find that for all protocol durations (both long enough to be in the adiabatic regime, as well as so short as to be in the non-adiabatic regime), the fidelity for $\omega\!\neq\! 0$ is lower than that for $\omega \!=\! 0$. This supports the claim that the bound to the speedup holds also when a non-commuting meter Hamiltonian is present.

In summary, we have shown that in the adiabatic regime with $[X_M, H_M]=0$, the maximum speedup achievable with our protocol is $1+m_{max}$, where $m_{max}$ is the maximal eigenvalue of $X_M$, the meter part of the system-meter interaction, and we have provided numerical evidence that this speedup bound holds even when both assumptions of adiabaticity and commutativity are relaxed.

\section{Application to quantum annealing}
\label{sec:protocol}

\subsection{Time-to-solution benchmark}
Time-to-solution is a widely used figure of merit in solving optimization problems with quantum annealing because it addresses the trade-off between annealing duration and success probability. Since solutions to combinatorial optimization problems are usually hard to find, but easy to verify, an algorithm that outputs the correct solution with any finite probability is sufficient, as it can then be repeated until the correct solution is obtained at least once. Therefore, one can either run the algorithm once with a large annealing duration $T$ and high probability of obtaining the correct result after one run, $p_{\text{single}}(T)$, or run it many times for smaller values of $T$, as well as anything in between the two extremes. Then, the time-to-solution (TTS), $\mathcal{T}$, is defined as
\begin{equation}
\label{eq:tts}
    \mathcal{T}_p = \min_{T} \; T \frac{\log(1-p)}{\log(1-p_{\text{single}}(T))}\,,
\end{equation}
and quantifies the expected \textit{total} time required to obtain the correct solution (or a particular correct solution when degeneracy is present) with probability $p$, minimized over different values of $T$. In what follows we set $p \!=\! 0.95$ without loss of generality.

\begin{figure}[tbp]
    (a) \\[-2ex]
    \includegraphics[width=0.45\textwidth]{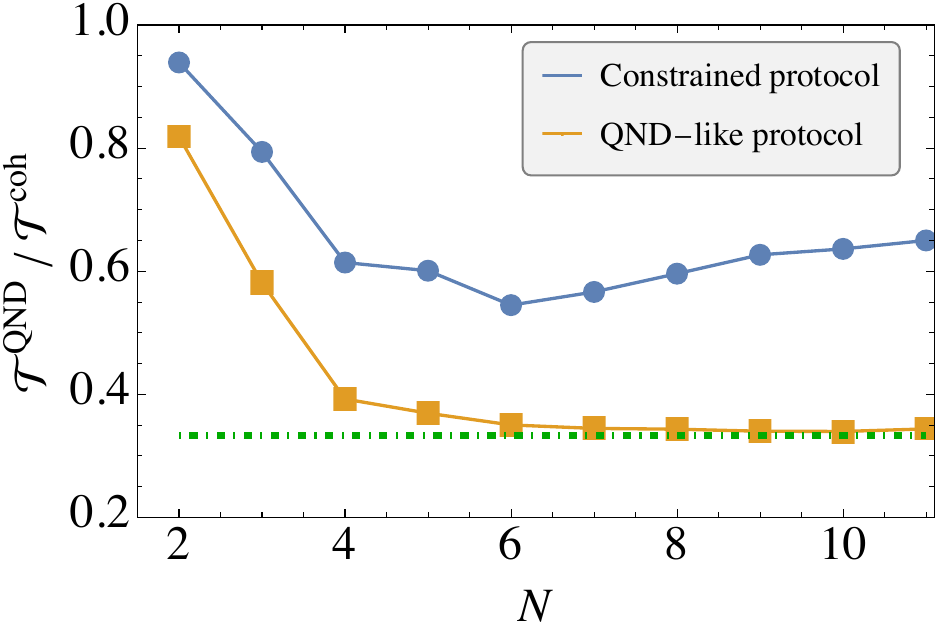}\\
    (b) \hskip0.45\columnwidth (c)\\
    \includegraphics[width=0.23\textwidth]{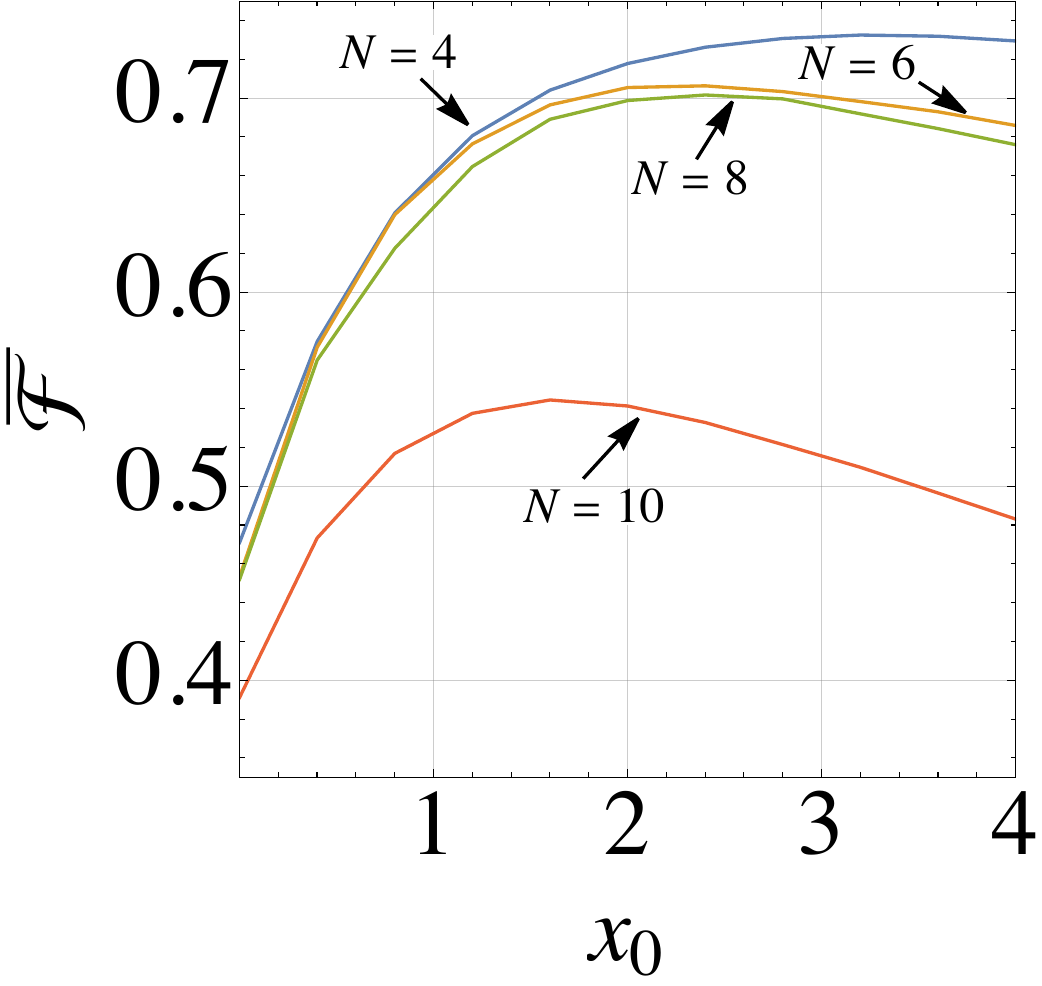} \includegraphics[width=0.23\textwidth]{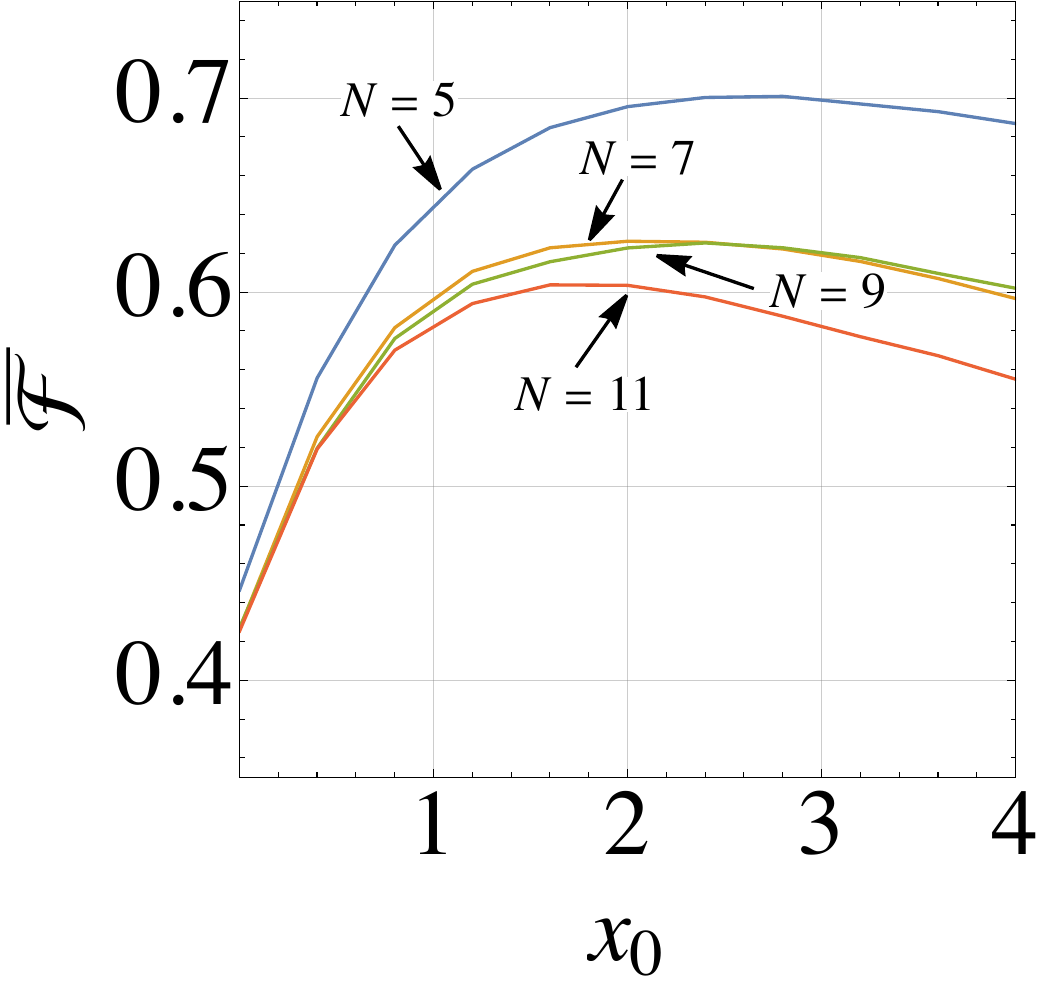}\\
    \caption{
    (a) Averaged relative speedup in time-to-solution, Eq.~\eqref{eq:TTS}, as a function of qubit number. For each problem size, the QND-like and constrained QND-like protocols are used to find ground states of $n\!=\! 100$ instances of random Ising Hamiltonians and we take the coupling strength with the meter to be $x_0\!=\!2$. The QND-like protocol reaches the predicted TTS ratio of $1/(1+x_0)=0.33$ (dashed green line). (b,c) Average fidelity  of the constrained QND-like protocol, as a function of system-meter interaction strength $x_0$, for different problem sizes (numbers of qubits), where we show even sizes in (b) and odd sizes in (c). For every problem size (number of system qubits), we run the protocol for $n\!=\! 100$ different instances of random Ising Hamiltonians, and compute the average fidelity. In both panels the meter has $X_M = x_0 \sigma_z$ and $H_M = 0$.}
    \label{fig:x0scan}
\end{figure}

We expect that the TTS speedup, defined as the ratio of the TTS for the QND-like protocol and the TTS for the coherent protocol, should scale exactly the same as for a single annealing schedule, discussed in Sec.~\ref{sec:speedup}, i.e. it should be equal to $1+m$, where $X_M\ket{m}=m\ket{m}$, and $\ket{m}$ is the initial meter state. This is because $p^{QND}_{single}(T) = p^{coherent}_{single}((1+m)T)$, as demonstrated in Sec.~\ref{sec:speedup}, therefore 
\begin{eqnarray}
\label{eq:TTS}
    \frac{\mathcal{T}_p^\text{QND}}{\mathcal{T}_p^\text{coh}} &= & \frac{\min_{T} T / \log(1-p_{single}(T))}{\min_{T} T / \log(1-p_{single}((1+m)T)))} \\
    & = & \frac{1}{1+m} \frac{\min_{T} T / \log(1-p_{single}(T))}{\min_{T'} T' / \log(1-p_{single}(T'))} \nonumber \\ 
    & = & \frac{1}{1+m} \nonumber,
\end{eqnarray}
where $T' \!=\! (1+m)T$. We verify this result numerically in Fig.~\ref{fig:x0scan}(a). We fix $H_S(t)$ from Eq.~\eqref{eq:H_QA} with $H_f$ set to be an Ising Hamiltonian, Eq.~\eqref{eqn:Ising}, with uniformly random parameters $J_{ij}\!\in\!\left[0,1\right]$ and $h_i\!\in\!\left[0,1\right]$. We also set $H_M \!=\! 0$, $X_M \!=\! x_0 \sigma_z$, and $\rho_M(0) \!=\! \ket{0}\bra{0}$ and therefore from Eq.~\eqref{eq:TTS} expect a speed up of $1/(1+x_0)$. We use $x_0 = 2.0$ which corresponds to the near optimal interaction strength value for a QND-like protocol where the allowed interactions are constrained such that difficult to realise terms are omitted, as discussed in the following subsection \ref{subsecQNDinspired} and whose performance is also shown in Figure~\ref{fig:x0scan}(a). We calculate the TTS ratio for $n \!=\! 100$ instances of random Ising Hamiltonians, plotting their average as a function of problem size, which is defined by the number of qubits, $N$, in the system. For the minimization over $T$ that is required in Eq.~\eqref{eq:tts} we use $n_T \!=\! 10$ values of $T$ which we determine using the method outlined in Appendix~\ref{AppendixA}. We observe that for $N\! >\! 3$ qubits, the speedup matches the derived result $1/(1+x_0) \!=\! 0.33$ and we remark that \chng{for $N<4$ the observed deviations are likely due to the finite sampling of $T$ in the minimization. For small systems, the higher occurrence of multiple avoided crossings with equal energy gap in the spectrum can lead to a poor choice for the annealing duration determined in Appendix~\ref{AppendixA}}.

\subsection{Constrained QND-like protocol}
\label{subsecQNDinspired}
If we consider the types of interactions needed to implement our protocol, Eq.~\eqref{eq:mainHamiltonian}, for solving QUBO problems as in the previous subsection, we find that for $X_M \!=\! \sigma_z$, one has to implement Hamiltonian terms $\sigma_x \sigma_z$, $\sigma_z \sigma_z$ and $\sigma_z \sigma_z \sigma_z$ in a simultaneous and time-dependent manner. Here we ask if one can recover the performance of the QND-like protocol with a more limited set of physical interactions.

In the context of quantum annealing, the system Hamiltonian $H_S(t)$ that appears in our protocol has the form
\begin{equation}
    H_S(t) = (1-f(t))H_i + f(t)H_f
\end{equation}
such that the ground state of $H_f$ encodes the problem solution. The constrained QND-like protocol is obtained by picking an interaction Hamiltonian $H_{int}(t) \!=\! f(t)H_f \otimes X_M$ instead of $H_{int}(t) = H_S(t) \otimes X_M$, such that only the problem Hamiltonian is coupled to the meter. As a result, fewer types of interactions need to be implemented. For example, in the aforementioned QUBO case, one would ``only" need to implement $\sigma_z \sigma_z$ and $\sigma_z \sigma_z \sigma_z$ terms, omitting the $\sigma_x \sigma_z$ term. The simplification comes at the cost that the interaction does not commute with the system Hamiltonian and therefore induces relaxation in addition to the energy rescaling. In Appendix~\ref{AppendixB} we furthermore consider decomposing the $\sigma_z \sigma_z \sigma_z$ term, as proposed in Ref.~\cite{decomposition}; however, such a decomposition comes with a quadratic overhead in number of ancilla qubits. While there have been some proposals for engineering three-body interactions in quantum computers and quantum simulators~\cite{ashkarin2022toffoli, fedorov2012implementation}, such interactions are typically weak. As a result, only a modest speedup can be expected, due to the linear dependence between speedup and interaction strength.

Since the unintended relaxation results in a more complex dynamics, we restrict to investigating the impact of the constrained protocol numerically. First, we study how its performance depends on the interaction strength, $x_0$. We again consider $n \!=\! 100$ instances of random Ising Hamiltonians for each problem size and plot the average fidelity, $\overline{\mathcal{F}}$, as a function of $x_0$ in Fig.~\ref{fig:x0scan}(b,c). We pick the annealing duration for each random Ising Hamiltonian instance separately again employing the method outlined in Appendix~\ref{AppendixA}. The data suggests that the fidelity improvement plateaus (importantly, not to 1) as a function of interaction strength at $x_0 \!\approx\! 2.0$, for all problem sizes considered~\footnote{\chng{We note that this optimal value of interaction strength is specific to our choice of parameters and expect that changes in e.g. the time-to-solution cutoff and/or choice of interactions to include in the constrained protocol, will have a non-trivial impact on this value.}}. This is in stark contrast to the full QND-like protocol, where a close to unit fidelity can always be reached by ramping up the interaction strength with the observed speedup scaling linearly. We suspect that the discrepancy arises because in the constrained QND-like protocol the unwanted relaxation becomes stronger in tandem with the desired energy rescaling effects as $x_0$ is increased.

Having established how the speedup depends on the interaction strength, we fix $x_0\! = \! 2.0$ and, just as for the ideal QND-like protocol, evaluate the TTS ratio averaged over $n \!=\! 100$ instances as a function of problem size. The result of the simulations is shown by the top-most (blue) curve in Figure~\ref{fig:x0scan}(a). This suggests that the contrained QND-like protocol has a significantly more limited performance, since for the ideal QND-like protocol, the speedup could be enhanced by an indefinite amount by ramping up the interaction strength; this appears not to be the case for the constrained QND-like protocol.

\subsection{Comparison with other control strategies}\label{STAcomparisons}
The basic mechanism by which the QND-like protocols achieve a speedup shares some qualitative similarity with other control techniques. As shown explicitly for the LZ model in Fig.~\ref{fig:coherences}(b), counterdiabatic driving achieves a speedup by changing the energy gaps in system; however, it does so in a very different manner to the protocol discussed here. While counterdiabatic approaches can be effective on, in principle, arbitrarily short time scales, this comes at the expense of needing to know the precise eigenstates of the controlled system. Moreover, it can lead to significant energetic costs in implementing the control~\cite{AbahNJP, CampbellEPL2023, CarolanImpulseControl}. Evidently then, this renders the direct application of such STA techniques ill-suited for QA problems. Recently, there have been efforts to alleviate the requirement of knowing the (time-dependent) eigenstates by exploiting a variational approach~\cite{SelsPNAS} whose applicability can be further augmented with the use of optimal control~\cite{delCampoPRA2014, cold} or Floquet engineering techniques~\cite{ClaeysPRL}. This has opened the possibility to leverage counterdiabatic techniques for QA problems, showing some significant promise~\cite{staQA1, staQA2, staQA3, passarelli2022optimal, passarelli2023counterdiabatic, Duncan2025}. \chng{It remains to be understood how such counterdiabatic techniques perform in QUBO-type problems where the coupling strengths in Eq.~\eqref{eq:H_QA} are random. While for uniform couplings, significant enhancements toward adiabaticity can be achieved even in the case of constrained control terms~\cite{cold, Duncan2025}, for more complex settings the relevance of difficult to realise $N$-body control terms tends to grow~\cite{Lawrence2025}. Thus, a benefit of our protocol is that it can be achieved with access to two- and three-body interaction terms only, although this necessitates the introduction of an additional auxiliary system (the meter).}

The QND-inspired protocol we propose is also complementary to these STA-type approaches in another important manner. Our protocol is based on an effective open system dynamics, whereas the STA approaches mentioned previously generally assume a closed unitary dynamics. Some STA schemes have been implemented with an auxilary system playing the role of a tunable environment~\cite{CarolanPRAGates, TouilEASTA}, however, these protocols require entanglement to be generated between the system and environment, in contrast to the protocol proposed here. 
A benefit of our protocol is the relative simplicity of the formulation, where no dynamical optimisation of the coupling is performed. 
The particular choice of which approach to take, i.e., directly manipulate the system via counterdiabatic techniques versus suitably engineer a dissipative dynamics or some combination of both, will naturally depend on the specific architecture in question which ultimately dictates the availability of the necessary time-dependent interaction terms.

\section{Discussion and conclusions}
\label{sec:concl}
We have investigated the speedups arising from, in principle decohering, QND-like annealing protocols and have demonstrated that, surprisingly, they arise from a coherent energy rescaling mechanism. By leveraging the commuting property that QND-like protocols possess, we have been able to derive, analytically and without any approximations, the speedup compared to the coherent annealing protocol. 
Our approach is based on an exactly solvable model which, while simple, might provide guidance for other noise engineering protocols.

We have also established exact bounds for the achievable speedup of our protocol for quantum annealing use cases. We have found that the speedup over the coherent protocol scales linearly with the interaction strength between meter and system, and does not scale with the problem size as captured by the number of qubits. On the one hand, this means that the speedup does not plateau as a function of the interaction strength. On the other hand, since the speedup stays constant as a function of problem size, it does not change the computational complexity of the annealing problem -  this is likely to be true for any energy rescaling based quantum annealing protocols. \chng{It is also worth noting that energy rescaling via coupling to an auxiliary degree of freedom can also provide speed ups for solving the systems of linear equations via adiabatic quantum computing methods~\cite{SommaPRL}. This suggests that our analysis, in particular in view of simplifying the required couplings between qubits and auxiliary degrees from freedom, may be relevant beyond QUBO problems. Moreover, it points to the fact that spectral gap amplification~\cite{Somma2013} is a simple but effective strategy that deserves to be explored more generally.}

Finally, from a practical, implementation-oriented point of view, we have briefly discussed the experimental feasibility of implementing our protocol while also alleviating the experimental difficulties associated with it, that is, the independent and time-dependent control of multiple $\sigma_z \sigma_z$, $\sigma_z \sigma_x$ and $\sigma_z \sigma_z \sigma_z$ terms. We have analysed a constrained QND-like protocol which removes the necessity for the $\sigma_z \sigma_x$ term, indicating that there is a trade-off between performance and ease of implementation. 

Our protocol is effective precisely because the coupling provides additional energy to the system \chng{and exactly mimics the dynamics that would be obtained by a direct energy rescaling of the original problem Hamiltonian in Eq.~\eqref{eq:H_QA}}. In future work, it would therefore be interesting to examine how such QND-like protocols perform when the maximum energy bandwidth is constrained to remain within a given window. It will also be interesting to understand precisely what are the physical mechanisms that allow various protocols achieve a speed up. 
Our protocol provides a versatile starting point to address this issue, one in which we can consider e.g. more complex, possibly many-body, meters as well as various system-meter couplings, allowing to establish the most relevant features that the control protocol must admit in order for finite time adiabatic dynamics to be realised.

\begin{acknowledgments}
Financial support from the German Federal Ministry of Education and Research (BMBF) within the project NiQ (13N16201) is gratefully acknowledged. SC acknowledges support from the Alexander von Humboldt Foundation and is grateful to Viv Kendon for useful discussions.
\end{acknowledgments}

\appendix
\section*{Appendices}
\section{Derivation of Eq.~(\ref{eq:corr1})}
\label{Appendix1}
Here we demonstrate how to express Eq.~\eqref{eqn:rho} in the instantaneous eigenbasis of $H_S(t)$. For this, we first consider pure states and start from the Schr\"odinger equation for the amplitudes in the instantaneous eigenbasis of $(1 \pm x_0) H_S(t)$, which is evidently the same eigenbasis as $H_S(t)$. With $\ket{\psi(t)} \!=\! \sum_m c_m(t) \ket{m(t)}$, the amplitudes evolve according to
\begin{eqnarray}
    \frac{d}{dt} c_m^{[1\pm x_0]}(t) &=& -i(1 \pm x_0)E_m(t)  c_m^{[1\pm x_0]}(t) \\ \nonumber &&- \braket{m(t)|\dot{m}(t)}  c_m^{[1\pm x_0]}(t) \\ &&+ \sum_{n \neq m} \frac{(\braket{m(t)|(1 \pm x_0) \dot{H_S(t)} | n(t)}}{(1 \pm x_0)(E_m(t) - E_n(t))} c_n^{[1\pm x_0]}(t)\,.\nonumber
\end{eqnarray}
Since $(1 \pm x_0)$ cancels in the last term, it only appears in the first (phase) term. 
We can use this to get a differential equation for the density matrices via
\begin{widetext}
\begin{eqnarray}
\label{eqn:rho_de}
    \frac{d}{dt} ((\rho^{[1 \pm x_0]}_S(t))_{mn}) &=& \frac{d}{dt} ((c_m^{[1\pm x_0]}(t))^*c_n^{[1\pm x_0]}(t)) 
  =   c_n^{[1\pm x_0]}(t) \frac{d}{dt}(c_m^{[1\pm x_0]}(t))^* + (c_m^{[1\pm x_0]}(t))^* \frac{d}{dt}(c_n^{[1\pm x_0]}(t))\\ &=& \nonumber 
     i(1 \pm x_0)(E_m(t) - E_n(t) (\rho^{[1 \pm x_0]}_S(t))_{mn} - (\braket{m(t)|\dot{m(t)}} - \braket{n(t)|\dot{n(t)}})(\rho^{[1 \pm x_0]}_S(t))_{mn} \\ &&+ \sum_{i \neq m}{\frac{(\braket{i(t)| \dot{H_S(t)} | m(t)}}{E_m(t) - E_i(t)}} (\rho^{[1 \pm x_0]}_S(t))_{mi} + \sum_{j \neq n}{\frac{(\braket{n(t)| \dot{H_S(t)} | j(t)}}{E_n(t) - E_j(t)}} (\rho^{[1 \pm x_0]}_S(t))_{jn}\,.\nonumber
\end{eqnarray}
Inserting Eq.~\eqref{eqn:rho_de} into Eq.~\eqref{eqn:rho} and simplifying, we obtain 
\begin{eqnarray}
    \frac{d}{dt} ((\rho_S(t))_{mn}) &= & i(E_m(t) - E_n(t)) \frac{1}{2}((\rho^{[1 + x_0]}_S(t))_{mn} + (\rho^{[1 - x_0]}_S(t))_{mn})\nonumber\\
    &&- (\braket{m(t)|\dot{m(t)}} - \braket{n(t)|\dot{n(t)}})\frac{1}{2}((\rho^{[1 + x_0]}_S(t))_{mn} + (\rho^{[1 - x_0]}_S(t))_{mn}) \nonumber \\
    && + \sum_{i \neq m}{\frac{(\braket{i(t)| \dot{H_S(t)} | m(t)}}{E_m(t) - E_i(t)}} \frac{1}{2}((\rho^{[1 + x_0]}_S(t))_{mi} + (\rho^{[1 - x_0]}_S(t))_{mi})\nonumber \\
    &&+ \sum_{j \neq n}{\frac{(\braket{n(t)| \dot{H_S(t)} | j(t)}}{E_n(t) - E_j(t)}} \frac{1}{2}((\rho^{[1 + x_0]}_S(t))_{jn} + (\rho^{[1 - x_0]}_S(t))_{jn}) \nonumber \\
    && + \frac{i x_0}{2}(E_m(t) - E_n(t))((\rho^{[1 + x_0]}_S(t))_{mn} - (\rho^{[1 - x_0]}_S(t))_{mn}).
\end{eqnarray}
The first four lines exactly match the rate of change of a density matrix evolving under $H_S(t)$, with the last line as a correction:
\begin{eqnarray}\label{eq:corr}
    \frac{d}{dt} \left(\rho_S(t)\right)_{mn} &=& -i\left(\left[H_S(t), \rho_S(t)\right]\right)_{mn} 
    + \frac{i x_0}{2} \left(E_m(t) - E_n(t)\right)\left((\rho^{[1 + x_0]}_S(t))_{mn} - (\rho^{[1 - x_0]}_S(t))_{mn}\right)\,.
\end{eqnarray}
\end{widetext}

\section{Choosing the annealing duration}
\label{AppendixA}
In this appendix, we outline how we pick the annealing duration for the parameter scan (Fig.~\ref{fig:x0scan}(b,c)) and time-to-solution scan (Fig. \ref{fig:x0scan}(a)). In both cases, the first step is to determine an annealing duration that would lead to an \textit{approximately} 50\% success probability. This is done with a two step process. First, the protocol is run for a guess duration $T_{guess}$ and the success probability $p$ is recorded. Then, we perform a crude extrapolation for what the duration should be to instead obtain $p=0.5$, based on the Landau-Zener tunneling formula, under which the infidelity decays exponentially with annealing time: $1-p \sim e^{-T}$. Thus, the extrapolated annealing time $T_{ext}$ to reach success probability of 50\% with a coherent protocol is
\begin{equation}
    T_{ext} = T_{guess} \frac{\log(1-p_{guess})}{\log(1-0.5)}.
\end{equation}
For the interaction strength scan, once $T_{ext}$ is obtained, we simply use this duration for different values of the interaction strength $x_0$. For the time-to-solution scan, $T_{ext}$ acts as the middle value (on a log scale) for all the annealing duration values $T$ used for the minimization over $T$: we pick $n = 10$ different values evenly spaced on a log scale, from $0.1T$ to $10T$.
The Landau-Zener tunneling formula used for the extrapolation is strictly only true if there is a single avoided crossing. However, practically speaking, achieving \textit{exactly} 50\% success probability is not essential for either the interaction strength scan or the time-to-solution scan.

\section{Decomposing 3-body Hamiltonian terms}
\label{AppendixB}
While the QND-like protocol introduced in Sec.~\ref{sec:protocol} produces a speedup compared to the coherent protocol, it (as well as the constrained QND-like protocol) requires individually tunable three-body $\sigma_z \sigma_z \sigma_z$ couplings, which are not widely accessible among present-day quantum devices. In this section, we discuss a possible implementation of the these protocols that utilise a decomposition of the three-body terms into experimentally accessible $\sigma_z \sigma_z$ and $\sigma_z$ terms, introduced in Ref.~\cite{decomposition}. 

The proposal recreates the ground state manifold of the $\sigma_z\sigma_z\sigma_z$ interaction, by connecting all 3 qubits participating in the interaction with a fourth ancilla qubit, 
\begin{eqnarray}
    \sigma_z^1 \sigma_z^2 \sigma_z^3 &\simeq& \sigma_z^1 \sigma_z^2 + \sigma_z^2 \sigma_z^3 + \sigma_z^1 \sigma_z^3 \\&&- \sum_{i=\{1,2,3\}} [2\sigma_z^i \sigma_z^a - \sigma_z^i] - 2\sigma_z^a\,,\nonumber
\end{eqnarray}
The $\simeq$ sign is to be understood as follows: for each of the four linearly independent ground states $\ket{\psi}$ of the left-hand side of the equation ($\ket{001}$, $\ket{010}$, $\ket{100}$, and $\ket{111}$), there exists a corresponding ground state $\ket{\phi}$ of the right-hand side of the equation, such that $\ket{\psi}\bra{\psi} = \text{Tr}_a \ket{\phi}\bra{\phi}$, where the trace is over the ancilla qubit. When the final state is measured, the annealing problem solution is obtained simply by tracing out (in this case just ignoring) the ancilla qubit. Furthermore, the gap to the first excited states is the same for both sides of the equation, this means that the decomposition should not increase the rate of diabatic transitions out of the ground state.

While the decomposition of the three-body interaction is accessible to current quantum devices without additional experimental improvements, it imposes a qubit overhead on the system due to the ancilla qubits needed. The number of extra qubits is equal to the number of non-zero $J_{ij}$ two-body couplings in the QUBO Hamiltonian; if the ``connectivity graph" of the QUBO problem is fully-connected (i.e. every $J_{ij}$ term is non-zero), then the qubit overhead grows quadratically with the number of logical qubits.

We believe such an overhead is too restrictive for the dissipative protocol to scale well in its ``decomposed" form. We see more promise in experimentally achieving the necessary 3-body interactions for the problem. Note that the required 3-body coupling need not be implemented for every triplet of qubits - one of the three bodies in the triplet is always the meter; instead of a qubit, it could be a motional mode in an ion trap or an optical mode in a cavity with neutral atoms.


\bibliographystyle{quantum}
\bibliography{bibliography}

\begin{thebibliography}{10}

\bibitem{Arimondo1996}
E.~Arimondo.
\newblock ``V {{Coherent Population Trapping}} in {{Laser Spectroscopy}}''.
\newblock In E.~Wolf, editor, Progress in {{Optics}}.
\newblock \href{https://dx.doi.org/10.1016/S0079-6638(08)70531-6}{Volume~35, pages 257--354}.
\newblock Elsevier~(1996).

\bibitem{ciraczollerreservoir1996}
J.~F. Poyatos, J.~I. Cirac, and P.~Zoller.
\newblock ``Quantum reservoir engineering with laser cooled trapped ions''.
\newblock \href{https://dx.doi.org/10.1103/PhysRevLett.77.4728}{Phys. Rev. Lett. {\bf 77}, 4728--4731}~(1996).

\bibitem{kraus2008reservoir}
B.~Kraus, H.~P. B\"uchler, S.~Diehl, A.~Kantian, A.~Micheli, and P.~Zoller.
\newblock ``Preparation of entangled states by quantum markov processes''.
\newblock \href{https://dx.doi.org/10.1103/PhysRevA.78.042307}{Phys. Rev. A {\bf 78}, 042307}~(2008).

\bibitem{diehl2008reservoir}
S.~Diehl, A.~Micheli, A.~Kantian, B.~Kraus, H.~P. B{\"u}chler, and P.~Zoller.
\newblock ``Quantum states and phases in driven open quantum systems with cold atoms''.
\newblock \href{https://dx.doi.org/10.1038/nphys1073}{Nature Physics {\bf 4}, 878--883}~(2008).

\bibitem{kienzler2015reservoir}
D.~Kienzler, H.-Y. Lo, B.~Keitch, L.~De~Clercq, F.~Leupold, F.~Lindenfelser, M.~Marinelli, V.~Negnevitsky, and J.~P. Home.
\newblock ``Quantum harmonic oscillator state synthesis by reservoir engineering''.
\newblock \href{https://dx.doi.org/10.1126/science.1261033}{Science {\bf 347}, 53--56}~(2015).

\bibitem{Verstraete:2008}
F.~Verstraete, M.~M. Wolf, and J.~I.~Cirac.
\newblock ``Quantum computation and quantum-state engineering driven by dissipation''.
\newblock \href{https://dx.doi.org/10.1038/nphys1342}{Nature Physics {\bf 5}, 633--636}~(2009).

\bibitem{pielawa2007generation}
S.~Pielawa, G.~Morigi, D.~Vitali, and L.~Davidovich.
\newblock ``Generation of einstein-podolsky-rosen-entangled radiation through an atomic reservoir''.
\newblock \href{https://dx.doi.org/10.1103/PhysRevLett.98.240401}{Phys. Rev. Lett. {\bf 98}, 240401}~(2007).

\bibitem{BurgarthPRL2007}
D.~Burgarth and V.~Giovannetti.
\newblock ``Full control by locally induced relaxation''.
\newblock \href{https://dx.doi.org/10.1103/PhysRevLett.99.100501}{Phys. Rev. Lett. {\bf 99}, 100501}~(2007).

\bibitem{raghunandan2020initialization}
M.~Raghunandan, F.~Wolf, C.~Ospelkaus, P.~O. Schmidt, and H.~Weimer.
\newblock ``Initialization of quantum simulators by sympathetic cooling''.
\newblock \href{https://dx.doi.org/10.1126/sciadv.aaw9268}{Science Advances {\bf 6}, eaaw9268}~(2020).

\bibitem{CampbellEPL}
S.~Campbell and B.~Vacchini.
\newblock ``Collision models in open system dynamics: A versatile tool for deeper insights?''.
\newblock \href{https://dx.doi.org/10.1209/0295-5075/133/60001}{EPL {\bf 133}, 60001}~(2021).

\bibitem{CiccarelloReview}
F.~Ciccarello, S.~Lorenzo, V.~Giovannetti, and G.~M. Palma.
\newblock ``Quantum collision models: Open system dynamics from repeated interactions''.
\newblock \href{https://dx.doi.org/10.1016/j.physrep.2022.01.001}{Physics Reports {\bf 954}, 1–70}~(2022).

\bibitem{raphael}
R.~Menu, J.~Langbehn, C.~P. Koch, and G.~Morigi.
\newblock ``Reservoir-engineering shortcuts to adiabaticity''.
\newblock \href{https://dx.doi.org/10.1103/PhysRevResearch.4.033005}{Phys. Rev. Res. {\bf 4}, 033005}~(2022).

\bibitem{King:2024}
E.~C. King, L.~Giannelli, R.~Menu, J.~N. Kriel, and G.~Morigi.
\newblock ``Adiabatic quantum trajectories in engineered reservoirs''.
\newblock \href{https://dx.doi.org/10.22331/q-2024-07-30-1428}{Quantum {\bf 8}, 1428}~(2024).

\bibitem{aqc}
T.~Albash and D.~A. Lidar.
\newblock ``Adiabatic quantum computation''.
\newblock \href{https://dx.doi.org/10.1103/RevModPhys.90.015002}{Rev. Mod. Phys. {\bf 90}, 015002}~(2018).

\bibitem{traffic}
F.~Neukart, G.~Compostella, C.~Seidel, D.~von Dollen, S.~Yarkoni, and B.~Parney.
\newblock ``Traffic flow optimization using a quantum annealer''.
\newblock \href{https://dx.doi.org/10.3389/fict.2017.00029}{Frontiers in ICT{\bf 4}}~(2017).

\bibitem{DeffnerTraffic2024}
K.~Domino, E.~Doucet, R.~Robertson, B.~Gardas, and S.~Deffner.
\newblock ``On the baltimore light raillink into the quantum future''~(2024).
\newblock  \href{http://arxiv.org/abs/2406.11268v1}{arXiv:2406.11268v1}.

\bibitem{scheduling}
D.~Venturelli, D.~J.~J. Marchand, and G.~Rojo.
\newblock ``Quantum annealing implementation of job-shop scheduling''~(2016).
\newblock  \href{http://arxiv.org/abs/1506.08479}{arXiv:1506.08479}.

\bibitem{portfolio_opt}
D.~Venturelli and A.~Kondratyev.
\newblock ``Reverse quantum annealing approach to portfolio optimization problems''.
\newblock \href{https://dx.doi.org/10.1007/s42484-019-00001-w}{Quantum Machine Intelligence {\bf 1}, 17--30}~(2019).

\bibitem{materials}
K.~Kitai, J.~Guo, S.~Ju, S.~Tanaka, K.~Tsuda, J.~Shiomi, and R.~Tamura.
\newblock ``Designing metamaterials with quantum annealing and factorization machines''.
\newblock \href{https://dx.doi.org/10.1103/PhysRevResearch.2.013319}{Phys. Rev. Res. {\bf 2}, 013319}~(2020).

\bibitem{phase_transitions}
R.~Harris, Y.~Sato, A.~J. Berkley, M.~Reis, F.~Altomare, M.~H. Amin, K.~Boothby, P.~Bunyk, C.~Deng, C.~Enderud, S.~Huang, E.~Hoskinson, M.~W. Johnson, E.~Ladizinsky, N.~Ladizinsky, T.~Lanting, R.~Li, T.~Medina, R.~Molavi, R.~Neufeld, T.~Oh, I.~Pavlov, I.~Perminov, G.~Poulin-Lamarre, C.~Rich, A.~Smirnov, L.~Swenson, N.~Tsai, M.~Volkmann, J.~Whittaker, and J.~Yao.
\newblock ``Phase transitions in a programmable quantum spin glass simulator''.
\newblock \href{https://dx.doi.org/10.1126/science.aat2025}{Science {\bf 361}, 162--165}~(2018).

\bibitem{topological}
A.~D. King, J.~Carrasquilla, J.~Raymond, I.~Ozfidan, E.~Andriyash, A.~Berkley, M.~Reis, T.~Lanting, R.~Harris, F.~Altomare, K.~Boothby, P.~I. Bunyk, C.~Enderud, A.~Fr{\'{e}}chette, E.~Hoskinson, N.~Ladizinsky, T.~Oh, G.~Poulin-Lamarre, C.~Rich, Y.~Sato, A.~Yu. Smirnov, L.~J. Swenson, M.~H. Volkmann, J.~Whittaker, J.~Yao, E.~Ladizinsky, M.~W. Johnson, J.~Hilton, and M.~H. Amin.
\newblock ``Observation of topological phenomena in a programmable lattice of 1,800 qubits''.
\newblock \href{https://dx.doi.org/10.1038/s41586-018-0410-x}{Nature {\bf 560}, 456--460}~(2018).

\bibitem{critical}
A.~D. King, J.~Raymond, T.~Lanting, R.~Harris, A.~Zucca, F.~Altomare, A.~J. Berkley, K.~Boothby, S.~Ejtemaee, C.~Enderud, E.~Hoskinson, S.~Huang, E.~Ladizinsky, A.~J.~R. MacDonald, G.~Marsden, R.~Molavi, T.~Oh, G.~Poulin-Lamarre, M.~Reis, C.~Rich, Y.~Sato, N.~Tsai, M.~Volkmann, J.~D. Whittaker, J.~Yao, A.~W. Sandvik, and M.~H. Amin.
\newblock ``Quantum critical dynamics in a 5,000-qubit programmable spin glass''.
\newblock \href{https://dx.doi.org/10.1038/s41586-023-05867-2}{Nature {\bf 617}, 61--66}~(2023).

\bibitem{scaling_advantage1}
T.~Albash and D.~A. Lidar.
\newblock ``Demonstration of a scaling advantage for a quantum annealer over simulated annealing''.
\newblock \href{https://dx.doi.org/10.1103/PhysRevX.8.031016}{Phys. Rev. X {\bf 8}, 031016}~(2018).

\bibitem{scaling_advantage2}
A.~D. King, J.~Raymond, T.~Lanting, S.~V. Isakov, M.~Mohseni, G.~Poulin-Lamarre, S.~Ejtemaee, W.~Bernoudy, I.~Ozfidan, A.~Yu. Smirnov, M.~Reis, F.~Altomare, M.~Babcock, C.~Baron, A.~J. Berkley, K.~Boothby, P.~I. Bunyk, H.~Christiani, C.~Enderud, B.~Evert, R.~Harris, E.~Hoskinson, S.~Huang, K.~Jooya, A.~Khodabandelou, N.~Ladizinsky, R.~Li, P.~A. Lott, A.~J.~R. MacDonald, D.~Marsden, G.~Marsden, T.~Medina, R.~Molavi, R.~Neufeld, M.~Norouzpour, T.~Oh, I.~Pavlov, I.~Perminov, T.~Prescott, C.~Rich, Y.~Sato, B.~Sheldan, G.~Sterling, L.~J. Swenson, N.~Tsai, M.~H. Volkmann, J.~D. Whittaker, W.~Wilkinson, J.~Yao, H.~Neven, J.~P. Hilton, E.~Ladizinsky, M.~W. Johnson, and M.~H. Amin.
\newblock ``Scaling advantage over path-integral monte carlo in quantum simulation of geometrically frustrated magnets''.
\newblock \href{https://dx.doi.org/10.1038/s41467-021-20901-5}{Nature Communications {\bf 12}, 1113}~(2021).

\bibitem{qa_review}
E.~J. Crosson and D.~A. Lidar.
\newblock ``Prospects for quantum enhancement with diabatic quantum annealing''.
\newblock \href{https://dx.doi.org/10.1038/s42254-021-00313-6}{Nature Reviews Physics {\bf 3}, 466--489}~(2021).

\bibitem{embedding_scaling}
M.~S. K\"onz, W.~Lechner, H.~G. Katzgraber, and M.~Troyer.
\newblock ``Embedding overhead scaling of optimization problems in quantum annealing''.
\newblock \href{https://dx.doi.org/10.1103/PRXQuantum.2.040322}{PRX Quantum {\bf 2}, 040322}~(2021).

\bibitem{santoro2002theory}
G.~E Santoro, R.~Marton{\'a}k, E.~Tosatti, and R.~Car.
\newblock ``Theory of quantum annealing of an ising spin glass''.
\newblock \href{https://dx.doi.org/10.1126/science.1068774}{Science {\bf 295}, 2427--2430}~(2002).

\bibitem{kadowaki1998quantum}
T.~Kadowaki and H.~Nishimori.
\newblock ``Quantum annealing in the transverse ising model''.
\newblock \href{https://dx.doi.org/10.1103/PhysRevE.58.5355}{Phys. Rev. E {\bf 58}, 5355--5363}~(1998).

\bibitem{passarelli2022optimal}
G.~Passarelli, R.~Fazio, and P.~Lucignano.
\newblock ``Optimal quantum annealing: A variational shortcut-to-adiabaticity approach''.
\newblock \href{https://dx.doi.org/10.1103/PhysRevA.105.022618}{Phys. Rev. A {\bf 105}, 022618}~(2022).

\bibitem{passarelli2023counterdiabatic}
G.~Passarelli and P.~Lucignano.
\newblock ``Counterdiabatic reverse annealing''.
\newblock \href{https://dx.doi.org/10.1103/PhysRevA.107.022607}{Phys. Rev. A {\bf 107}, 022607}~(2023).

\bibitem{seki2012quantum}
Y.~Seki and H.~Nishimori.
\newblock ``Quantum annealing with antiferromagnetic fluctuations''.
\newblock \href{https://dx.doi.org/10.1103/PhysRevE.85.051112}{Phys. Rev. E {\bf 85}, 051112}~(2012).

\bibitem{Najafabadi2023}
M.~S.~Najafabadi, D.~Schumayer, C.-K. Lee, D.~Jaksch, and D.~A.~W. Hutchinson.
\newblock ``Improving quantum annealing by engineering the coupling to the environment''.
\newblock \href{https://dx.doi.org/10.1140/epjqt/s40507-023-00202-0}{EPJ Quantum Technology {\bf 10}, 1--13}~(2023).

\bibitem{CuginiPRR2025}
D.~Cugini, D.~Nigro, M.~Bruno, and D.~Gerace.
\newblock ``Exponential optimization of adiabatic quantum-state preparation''.
\newblock \href{https://dx.doi.org/10.1103/PhysRevResearch.7.L012074}{Phys. Rev. Res. {\bf 7}, L012074}~(2025).

\bibitem{sta_review}
D.~Gu\'ery-Odelin, A.~Ruschhaupt, A.~Kiely, E.~Torrontegui, S.~Mart\'{\i}nez-Garaot, and J.~G. Muga.
\newblock ``Shortcuts to adiabaticity: Concepts, methods, and applications''.
\newblock \href{https://dx.doi.org/10.1103/RevModPhys.91.045001}{Rev. Mod. Phys. {\bf 91}, 045001}~(2019).

\bibitem{berry_cd}
M.~V. Berry.
\newblock ``Transitionless quantum driving''.
\newblock \href{https://dx.doi.org/10.1088/1751-8113/42/36/365303}{J. Phys. A {\bf 42}, 365303}~(2009).

\bibitem{delCampoPRA2014}
H.~Saberi, T.~Opatrn\'y, K.~M\o{}lmer, and A.~del Campo.
\newblock ``Adiabatic tracking of quantum many-body dynamics''.
\newblock \href{https://dx.doi.org/10.1103/PhysRevA.90.060301}{Phys. Rev. A {\bf 90}, 060301}~(2014).

\bibitem{cold}
I.~{\v{C}}epait{\.{e}}, A.~Polkovnikov, A.~J. Daley, and C.~W. Duncan.
\newblock ``Counterdiabatic optimized local driving''.
\newblock \href{https://dx.doi.org/10.1103/PRXQuantum.4.010312}{PRX Quantum {\bf 4}, 010312}~(2023).

\bibitem{invariant_engineering}
X.~Chen, I.~Lizuain, A.~Ruschhaupt, D.~Gu\'ery-Odelin, and J.~G. Muga.
\newblock ``Shortcut to adiabatic passage in two- and three-level atoms''.
\newblock \href{https://dx.doi.org/10.1103/PhysRevLett.105.123003}{Phys. Rev. Lett. {\bf 105}, 123003}~(2010).

\bibitem{Shingu2025}
Y.~Shingu and T.~Hatomura.
\newblock ``Geometrical scheduling of adiabatic control without information of energy spectra''~(2025).
\newblock  \href{http://arxiv.org/abs/2501.11846}{arXiv:2501.11846}.

\bibitem{Vacanti:2014}
G.~Vacanti, R.~Fazio, S.~Montangero, G.~M. Palma, M.~Paternostro, and V.~Vedral.
\newblock ``Transitionless quantum driving in open quantum systems''.
\newblock \href{https://dx.doi.org/10.1088/1367-2630/16/5/053017}{New J. Phys. {\bf 16}, 053017}~(2014).

\bibitem{ChenuPRR}
L.~Dupays, I.~L. Egusquiza, A.~del Campo, and A.~Chenu.
\newblock ``Superadiabatic thermalization of a quantum oscillator by engineered dephasing''.
\newblock \href{https://dx.doi.org/10.1103/PhysRevResearch.2.033178}{Phys. Rev. Res. {\bf 2}, 033178}~(2020).

\bibitem{avron2011landau}
J.~E. Avron, M.~Fraas, G.~M. Graf, and P.~Grech.
\newblock ``Landau-zener tunneling for dephasing lindblad evolutions''.
\newblock \href{https://dx.doi.org/10.1007/s00220-011-1269-y}{Communications in mathematical physics {\bf 305}, 633--639}~(2011).

\bibitem{haroche2006exploring}
S.~Haroche and J.-M. Raimond.
\newblock ``Exploring the quantum: atoms, cavities, and photons''.
\newblock \href{https://dx.doi.org/10.1093/acprof:oso/9780198509141.001.0001}{Oxford university press}. ~(2006).

\bibitem{glover2019quantum}
F.~Glover, G.~Kochenberger, and Y.~Du.
\newblock ``Quantum bridge analytics i: a tutorial on formulating and using qubo models''.
\newblock \href{https://dx.doi.org/10.1007/s10479-022-04634-2}{Ann. Oper. Res. {\bf 17}, 335--371}~(2019).

\bibitem{AutomaticQUBO}
D.~Volpe, N.~Quetschlich, M.~Graziano, G.~Turvani, and R.~Wille.
\newblock ``Towards an automatic framework for solving optimization problems with quantum computers''~(2024).
\newblock  \href{http://arxiv.org/abs/2406.12840}{arXiv:2406.12840}.

\bibitem{itano1990quantum}
W.~M. Itano, D.~J. Heinzen, J.~J. Bollinger, and D.~J. Wineland.
\newblock ``Quantum zeno effect''.
\newblock \href{https://dx.doi.org/10.1103/PhysRevA.41.2295}{Phys. Rev. A {\bf 41}, 2295--2300}~(1990).

\bibitem{AbahNJP}
O.~Abah, R.~Puebla, A.~Kiely, G.~De~Chiara, M.~Paternostro, and S.~Campbell.
\newblock ``Energetic cost of quantum control protocols''.
\newblock \href{https://dx.doi.org/10.1088/1367-2630/ab4c8c}{New J. Phys. {\bf 21}, 103048}~(2019).

\bibitem{roland2002quantum}
J.~Roland and N.~J. Cerf.
\newblock ``Quantum search by local adiabatic evolution''.
\newblock \href{https://dx.doi.org/10.1103/PhysRevA.65.042308}{Phys. Rev. A {\bf 65}, 042308}~(2002).

\bibitem{Landau1932}
L.~D. Landau.
\newblock ``{A theory of energy transfer. II}''.
\newblock \href{https://dx.doi.org/10.1016/B978-0-08-010586-4.50014-6}{Phys. Z. Sowjet {\bf 2}, 46}~(1932).

\bibitem{Zener1932}
C.~Zener.
\newblock ``{Non-adiabatic crossing of energy levels}''.
\newblock \href{https://dx.doi.org/10.1098/rspa.1932.0165}{Proc. R. Soc. A {\bf 33}, 696--702}~(1932).

\bibitem{GarrawayPRA}
N.~V. Vitanov and B.~M. Garraway.
\newblock ``Landau-zener model: Effects of finite coupling duration''.
\newblock \href{https://dx.doi.org/10.1103/PhysRevA.53.4288}{Phys. Rev. A {\bf 53}, 4288--4304}~(1996).

\bibitem{decomposition}
M.~Leib, P.~Zoller, and W.~Lechner.
\newblock ``A transmon quantum annealer: decomposing many-body ising constraints into pair interactions''.
\newblock \href{https://dx.doi.org/10.1088/2058-9565/1/1/015008}{Quantum Science and Technology {\bf 1}, 015008}~(2016).

\bibitem{ashkarin2022toffoli}
I.~N. Ashkarin, I.~I. Beterov, E.~A. Yakshina, D.~B. Tretyakov, V.~M. Entin, I.~I. Ryabtsev, P.~Cheinet, K.-L. Pham, S.~Lepoutre, and P.~Pillet.
\newblock ``Toffoli gate based on a three-body fine-structure-state-changing f\"orster resonance in rydberg atoms''.
\newblock \href{https://dx.doi.org/10.1103/PhysRevA.106.032601}{Phys. Rev. A {\bf 106}, 032601}~(2022).

\bibitem{fedorov2012implementation}
A.~Fedorov, L.~Steffen, M.~Baur, M.~P. da~Silva, and A.~Wallraff.
\newblock ``Implementation of a toffoli gate with superconducting circuits''.
\newblock \href{https://dx.doi.org/10.1038/nature10713}{Nature {\bf 481}, 170--172}~(2012).

\bibitem{CampbellEPL2023}
S.~Campbell.
\newblock ``Quantum work statistics of controlled evolutions''.
\newblock \href{https://dx.doi.org/10.1209/0295-5075/acfb33}{EPL {\bf 143}, 68001}~(2023).

\bibitem{CarolanImpulseControl}
E.~Carolan, A.~Kiely, and S.~Campbell.
\newblock ``Counterdiabatic control in the impulse regime''.
\newblock \href{https://dx.doi.org/10.1103/PhysRevA.105.012605}{Phys. Rev. A {\bf 105}, 012605}~(2022).

\bibitem{SelsPNAS}
D.~Sels and A.~Polkovnikov.
\newblock ``Minimizing irreversible losses in quantum systems by local counterdiabatic driving''.
\newblock \href{https://dx.doi.org/10.1073/pnas.1619826114}{Proc. Natl. Acad. Sci. {\bf 114}, E3909}~(2017).

\bibitem{ClaeysPRL}
P.~W. Claeys, M.~Pandey, D.~Sels, and A.~Polkovnikov.
\newblock ``Floquet-engineering counterdiabatic protocols in quantum many-body systems''.
\newblock \href{https://dx.doi.org/10.1103/PhysRevLett.123.090602}{Phys. Rev. Lett. {\bf 123}, 090602}~(2019).

\bibitem{staQA1}
K.~Takahashi.
\newblock ``Shortcuts to adiabaticity for quantum annealing''.
\newblock \href{https://dx.doi.org/10.1103/PhysRevA.95.012309}{Phys. Rev. A {\bf 95}, 012309}~(2017).

\bibitem{staQA2}
L.~Prielinger, A.~Hartmann, Y.~Yamashiro, K.~Nishimura, W.~Lechner, and H.~Nishimori.
\newblock ``Two-parameter counter-diabatic driving in quantum annealing''.
\newblock \href{https://dx.doi.org/10.1103/PhysRevResearch.3.013227}{Phys. Rev. Res. {\bf 3}, 013227}~(2021).

\bibitem{staQA3}
P.~R. Hegde, G.~Passarelli, A.~Scocco, and P.~Lucignano.
\newblock ``Genetic optimization of quantum annealing''.
\newblock \href{https://dx.doi.org/10.1103/PhysRevA.105.012612}{Phys. Rev. A {\bf 105}, 012612}~(2022).

\bibitem{Duncan2025}
C.~W. Duncan.
\newblock ``Counterdiabatic-influenced floquet-engineering: State preparation, annealing and learning the adiabatic gauge potential''~(2025).
\newblock  \href{http://arxiv.org/abs/2501.14881v1}{arXiv:2501.14881v1}.

\bibitem{Lawrence2025}
E.~Lawrence, S.~F.~J. Schmid, I.~{\v{C}}epait{\.{e}}, P.~Kirton, and C.~W. Duncan.
\newblock ``A numerical approach for calculating exact non-adiabatic terms in quantum dynamics''.
\newblock \href{https://dx.doi.org/10.21468/scipostphys.18.1.014}{SciPost Physics {\bf 18}, 014}~(2025).

\bibitem{CarolanPRAGates}
E.~Carolan, B.~\c{C}akmak, and S.~Campbell.
\newblock ``Robustness of controlled hamiltonian approaches to unitary quantum gates''.
\newblock \href{https://dx.doi.org/10.1103/PhysRevA.108.022423}{Phys. Rev. A {\bf 108}, 022423}~(2023).

\bibitem{TouilEASTA}
A.~Touil and S.~Deffner.
\newblock ``Environment-assisted shortcuts to adiabaticity''.
\newblock \href{https://dx.doi.org/10.3390/e23111479}{Entropy {\bf 23}, 1479}~(2021).

\bibitem{SommaPRL}
Y.~Suba\c{s}\i, R.~D. Somma, and D.~Orsucci.
\newblock ``Quantum algorithms for systems of linear equations inspired by adiabatic quantum computing''.
\newblock \href{https://dx.doi.org/10.1103/PhysRevLett.122.060504}{Phys. Rev. Lett. {\bf 122}, 060504}~(2019).

\bibitem{Somma2013}
R.~D. Somma and S.~Boixo.
\newblock ``Spectral gap amplification''.
\newblock \href{https://dx.doi.org/10.1137/120871997}{SIAM Journal on Computing {\bf 42}, 593–610}~(2013).

\end{thebibliography}

\end{document}